\documentclass[a4paper,11pt]{article}
\pdfoutput=1 

\usepackage{amssymb}
\usepackage{jheppub}
\usepackage{physics}
\usepackage{autobreak}
\usepackage{hyperref}
\usepackage{color}
\usepackage{natbib}
\usepackage{graphicx}
\hypersetup{colorlinks=true, citecolor=blue, urlcolor=blue, linkcolor=blue}

\title{\boldmath Tricritical phenomena in holographic chiral phase transitions}
\author{Masataka Matsumoto}
\affiliation{Department of Mathematics, Shanghai University,\\
	Shangda road 99, Shanghai 200444, China}
\emailAdd{matsumoto@shu.edu.cn}

\abstract{
	We study critical phenomena at a tricritical point associated with a chiral phase transition which emerges in the D3/D7 model in the presence of a finite baryon number density and an external magnetic field. 
	We numerically determine critical exponents related to the thermodynamic quantities and correlation functions. 
	We find that the values of the critical exponents agree with the mean-field values. 
	The scaling relations between the critical exponents are satisfied, implying that the scaling hypothesis for the free energy and the correlation functions hold. 
	Our results indicate that the critical phenomena at the tricritical point in the D3/D7 model are well described by the conventional Landau theory.}

\begin{document} 
	\maketitle
	\flushbottom
	
	\section{Introduction} \label{sec:intro}
	The AdS/CFT correspondence\,\cite{Maldacena:1998,Gubser:1998,Witten:1998} has been extensively used to study the thermodynamic and transport properties of strongly correlated quantum field theories at finite temperature.
	Applications of the correspondence to many areas of physics, such as quantum chromodynamics (QCD) \cite{Casalderrey-Solana:2011dxg,Kim:2012ey} and condensed matter physics \cite{zaanen_liu_sun_schalm_2015,Hartnoll:2016apf}, have been studied in the last two decades.
	In this paper, we focus particularly on critical phenomena at a tricritical point (TCP) that is one of universal phenomena in QCD and condensed matter physics.
	In QCD, there is a TCP associated with the chiral phase transition in the chiral limit \cite{Halasz:1998qr}.
	In condensed matter physics, on the other hand, a TCP can be observed, for example, in a metamagnet such as a FeCl$_{2}$ crystal \cite{PhysRev.164.866,PhysRevLett.31.1414}.
	
	In the context of the AdS/CFT correspondence, there are in general two different approaches to study the holographic dual description of QCD or condensed matter physics: a top-down approach and a bottom-up approach.
	For example, one of the QCD-like models in the top-down approach is the D3/D7 model \cite{Karch:2002sh}, where the D3-branes represent the gauge degrees of freedom and a probe D7-brane represents the flavor degrees of freedom in the dual field theory.
	To be precise, the D3/D7 model is dual to the ${\cal{N}}=4$ supersymmetric Yang-Mills theory with the ${\cal{N}}=2$ hypermultiplets.
	The dual system undergoes the chiral phase transition at finite temperature in the presence of the magnetic field and the baryon number density \cite{Evans:2010iy} (see also \cite{Filev:2007gb,Albash:2007bk}), although the transition is between the different classical solutions in the bulk picture.
	Considering the massless quark, the TCP associated with the chiral phase transition emerges\,\cite{Evans:2010iy} as in QCD.
	The corresponding Nambu-Goldstone modes have been studied in \cite{Filev:2009xp,Ishigaki:2020vtr}.
	In this paper, we investigate the critical phenomena at the TCP in the D3/D7 model.
	
	In general, critical phenomena are characterized by a few critical exponents associated with the critical behaviors of the physical quantities.
	The critical exponents are related to each other through the scaling relations which are justified by the scaling hypothesis in the vicinity of the critical point.
	Phenomenologically, the Landau theory presents an effective description of critical phenomena near a critical point.
	In the conventional Landau theory, critical exponents take the mean-field values and a system belongs to the mean-field universality class.
	However, it is not so obvious that the tricritical phenomena in the D3/D7 model obey the scaling theory and the Landau theory because the system is not a standard statistical system but a classical gravity system.
	In this paper, we numerically determine the values of the critical exponents and confirm if the scaling relations among them are satisfied at the TCP.
	Then, we determine the universality class of the TCP in the chiral phase transition of the D3/D7 model.
	This is the main purpose of this paper.
	The analysis of the tricritical phenomena in this paper will follow the discussion in \cite{Domb:1984}.
	
	One of the motivations of this paper is to reveal the difference between the tricritical phenomena in the equilibrium regime and the non-equilibrium regime.
	The author studied tricritical phenomena in the non-equilibrium steady state using the D3/D7 model \cite{Matsumoto:2022nqu}, where we employ the current density as a parameter.
	In \cite{Matsumoto:2022nqu}, we found that the tricritical phenomena is asymmetric with respect to the chiral symmetry restored\,($\chi$SR) phase and broken\,($\chi$SB) phase.
	Furthermore, the values of the critical exponents ($\gamma,\nu$) in the $\chi$SB phase are not the mean-field values, whereas those in the $\chi$SR phase are the mean-field values.
	Note that the other critical exponents determined in \cite{Matsumoto:2022nqu} agree with the mean-field values irrespective to the $\chi$SR phase or $\chi$SB phase.
	That is why we study the tricritical phenomena in the equilibrium regime and compare them to those in the non-equilibrium regime.

	The paper is organized as follows.
	In section \ref{sec:tcp}, we present a review of the scaling theory at a TCP.
	In section \ref{sec:Landau}, we discuss the tricritical behaviors in the framework of the Landau theory.
	In section \ref{sec:D3D7}, we introduce our holographic setup and present the main results of this paper, that is, the tricritical phenomena in the D3/D7 model.
	Section \ref{sec:conclusion} is devoted to conclusion and discussions.
	In the appendix, we discuss a method for the calculation of the Green's functions in the $\chi\text{SB}$ phase.
	
	\section{Scaling theory of tricritical points}
	\label{sec:tcp}
	In this section, we briefly review the scaling theory of TCPs. A critical point is defined as the end of a two-phase coexistence line. In analogy, a TCP is defined as the end of a three-phase coexistence line, the so-called triple line. The most essential feature of critical phenomena is a singular behavior of physical quantities, such as the order parameter and susceptibilities, in the vicinity of the critical point or the TCP. In general, the scaling theory is a strong tool for studying the critical behaviors. 
	
\begin{figure}[tbp]
		\centering 
		\includegraphics[width=12cm]{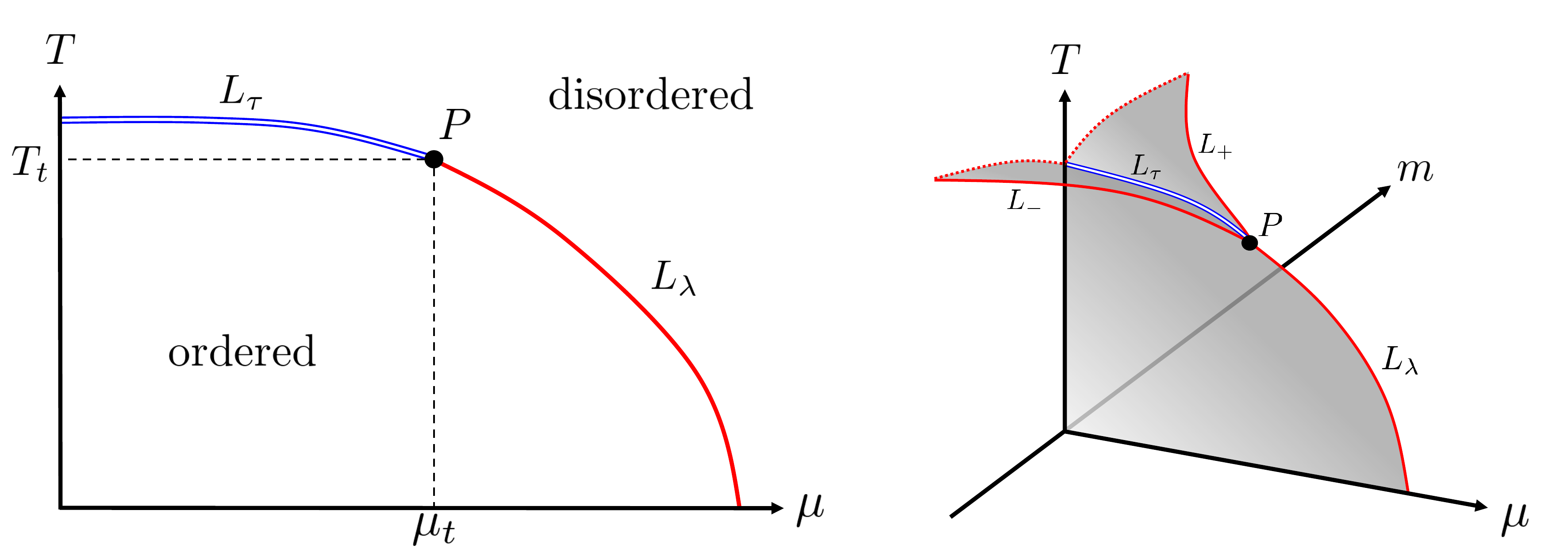}
		\caption{Schematic phase diagram of a system in which a TCP emerges. We use temperature $T$, chemical potential $\mu$, and quark mass $m$ as parameters. The left panel shows the phase diagram of the $m=0$ plane. The blue double and red solid line denote the first and second order phase transition points, called the triple line ($L_{\tau}$) and $\lambda$-line ($L_{
				\lambda}$), respectively. The point $P$ denotes the TCP. In the three dimensional phase diagram (right panel), two extra red lines denote the second order phase transition points, corresponding to the lines of the critical end points ($L_{\pm}$). The gray shaded surfaces denote the first-order phase transition points.}
		\label{fig:oekaki}
	\end{figure}

	We suppose a system in which a TCP emerges, such as a metamagnet or quark matter in the chiral limit.
	With a chiral symmetry breaking in mind, we consider the phase diagram as shown in Figure \ref{fig:oekaki}, where we employ three parameters: temperature $T$, chemical potential $\mu$, and quark mass $m$.
	Note that the quark mass $m$ is an external source that explicitly breaks the chiral symmetry. 
	In the plane of $m=0$ (left panel of Figure \ref{fig:oekaki}), there are lines of first- and second-order phase transitions, called the triple line ($L_{\tau}$ in Figure \ref{fig:oekaki}) and the $\lambda$-line ($L_{\lambda}$ in Figure \ref{fig:oekaki}), respectively.
	By definition, the point $P$ denotes the TCP.
	
	The $\lambda$-line ($T\leq T_{t}$) can generally be expanded near the TCP as
	\begin{equation}
		\mu_{\lambda}(T) = \mu_{t}- A_{0} \,\frac{T-T_{t}}{T_{t}} + \cdots,
	\end{equation}
	where $\mu_{t}$ and $T_{t}$ are the chemical potential and temperature at the TCP. The dots denote higher order terms in $T-T_{t}$. The triple line for $T\geq T_{t}$ can be written as
	\begin{equation}
		\mu_{\tau}(T) = \mu_{t} - A_{0} \,\frac{T-T_{t}}{T_{t}} + \cdots.
	\end{equation}
	Here, we assume that the $\lambda$-line and the triple line are asymptotically parallel at the TCP. Thus, a positive constant $A_{0}$ relates to the slope at the TCP. We introduce the scaling fields:
	\begin{equation}
		t=\frac{T-T_{t}}{T_{t}} , \hspace{0.5em} g=\left(\mu-\mu_{t} \right) + A_{0}t, \hspace{0.5em} h=m.
		\label{eq:scalingfield}
	\end{equation}
	Note that the TCP is at $t=g=h=0$.
	One can see that the path of $\abs{t}\to 0$ with $g=0$ is approaching the TCP and asymptotically tangential to the phase boundary.
	
	A tricritical scaling hypothesis can be formulated in terms of the scaling fields $t$, $g$, and $h$ \cite{Riedel1972a,Riedel1972b}. We assume that the singular part of the free energy in the vicinity of the TCP can be written as 
	\begin{equation}
		G_{\rm sing}(t,g,h) \simeq \abs{t}^{2-\alpha} {\cal{G}}^{(\pm)} \left( \frac{g}{\abs{t}^{\phi}}, \frac{h}{\abs{t}^{\Delta}} \right),
		\label{eq:Gscale}
	\end{equation}
	where ${\cal G}^{(\pm)}(x)$ is the scaling function which is assumed to be regular at the TCP. The superscript $(\pm)$ represents the sign of $t$. The exponents $(\alpha,\phi,\Delta)$ are the tricritical exponents. To be specific, $\alpha$ is called the thermal exponent and $(\phi,\Delta)$ are called the crossover exponents. In order to describe the critical behaviors near the TCP correctly, the scaling function ${\cal G}^{(\pm)}(x)$ must satisfy some requirements, which leads to some scaling relations between the critical exponents.
	Similarly, the scaling hypothesis can also be written as
	\begin{equation}
		G_{\rm sing}(t,g,h) \simeq \abs{g}^{2-\alpha_{t}} {\cal{G}}^{(\pm)}_{t} \left( \frac{t}{\abs{g}^{\phi_{t}}}, \frac{h}{\abs{g}^{\Delta_{t}}} \right),
	\end{equation}
	where the sets of exponents $(\alpha_{t},\phi_{t},\Delta_{t})$ are related to $(\alpha,\phi,\Delta)$ through
	\begin{equation}
		2-\alpha_{t} = \frac{2-\alpha}{\phi} , \hspace{0.5em} \phi_{t}=\frac{1}{\phi}, \hspace{0.5em} \Delta_{t} = \frac{\Delta}{\phi}.
	\end{equation}
	The exponents $(\alpha,\phi,\Delta)$ describe the leading singular behavior when we consider a path to the TCP that is tangential to the phase boundary, whereas $(\alpha_{t},\phi_{t},\Delta_{t})$ characterizes the singular behavior along a path to the TCP at a finite angle with the phase boundary.
	Hence, we distinguish between the critical behaviors along a path that is tangential and not tangential to the phase boundary.
	We will denote the subscript $t$ for the critical exponents in the latter case, that is, a path that is not tangential to the phase boundary.
	
	Let us see the scaling relations that should be satisfied near the TCP. 
	Here, we first consider a path tangentially approaching to the phase boundary.
	The order parameter is defined by
	\begin{equation}
		\sigma(t,g,h) = -\left( \frac{\partial G_{\rm sing}}{\partial h}\right)_{t,g},
	\end{equation}
	where the derivative is evaluated with other variables $(t,g)$ fixed. 
	Similarly, we define the ``density" by
	\begin{equation}
		\sigma_{2}(t,g,h) = -\left( \frac{\partial G_{\rm sing}}{\partial g}\right)_{t,h}.
	\end{equation}
	Since it is natural to assume that the scaling hypothesis can be applied to the derivative of the free energy, the order parameter and density can also be written as the following scaling forms near the TCP:
	\begin{equation}
		\sigma(t,g,h) \simeq  \abs{t}^{\beta} {\cal{M}}^{(\pm)} \left( \frac{g}{\abs{t}^{\phi}}, \frac{h}{\abs{t}^{\Delta}} \right), \hspace{0.5em}
		\sigma_{2}(t,g,h) \simeq  \abs{t}^{\beta_{2}} {\cal{M}}^{(\pm)}_{2} \left( \frac{g}{\abs{t}^{\phi}}, \frac{h}{\abs{t}^{\Delta}} \right),
		\label{eq:scalingOP}
	\end{equation}
	where ${\cal{M}}^{(\pm)}(x,y)$ and ${\cal{M}}^{(\pm)}_{2}(x,y)$ are given by
	\begin{equation}
		{\cal{M}}^{(\pm)}(x,y) = -\frac{\partial}{\partial y} {\cal{G}}^{(\pm)}(x,y), \hspace{0.5em}
		{\cal{M}}^{(\pm)}_{2}(x,y) = -\frac{\partial}{\partial x} {\cal{G}}^{(\pm)}(x,y).
	\end{equation}
	The exponents $\beta$ and $\beta_{2}$ are defined by the critical behavior of the order parameter and density in the vicinity of the TCP, respectively.
	The scaling forms of (\ref{eq:scalingOP}) imply the scaling relations:
	\begin{equation}
		\beta=2-\alpha-\Delta, \hspace{0.5em} \beta_{2} = 2-\alpha-\phi.
		\label{eq:scaling1}
	\end{equation}
	In addition, the corresponding susceptibilities are defined by
	\begin{equation}
		\chi(t,g,h) = \left(\frac{\partial \sigma}{\partial h} \right)_{t,g}, \hspace{0.5em} \chi_{2}(t,g,h) = \left(\frac{\partial \sigma_{2}}{\partial g} \right)_{t,h}.
	\end{equation}
	In analogy, the scaling forms of the susceptibilities can be written as
	\begin{equation}
		\chi(t,g,h) \simeq \abs{t}^{-\gamma} {\cal{X}}^{(\pm)}\left(\frac{g}{\abs{t}^{\phi}},\frac{h}{\abs{t}^{\Delta}} \right), \hspace{0.5em}
		\chi_{2}(t,g,h) \simeq \abs{t}^{-\gamma_{2}} {\cal{X}}^{(\pm)}_{2}\left(\frac{g}{\abs{t}^{\phi}},\frac{h}{\abs{t}^{\Delta}} \right),
		\label{eq:scalingS}
	\end{equation}
	where the scaling functions ${\cal{X}}^{(\pm)}(x,y)$ and ${\cal{X}}^{(\pm)}_{2}(x,y)$ are given by
	\begin{equation}
		{\cal{X}}^{(\pm)}(x,y) = -\frac{\partial^{2}}{\partial y^{2}}{\cal{G}}^{(\pm)}(x,y), \hspace{0.5em}
		{\cal{X}}^{(\pm)}_{2}(x,y) = -\frac{\partial^{2}}{\partial x^{2}}{\cal{G}}^{(\pm)}(x,y).
	\end{equation}
	Thus, $\gamma$ and $\gamma_{2}$ are defined by the critical behaviors of the corresponding susceptibilities.
	Similarly, we obtain the following scaling relations from (\ref{eq:scalingS})
	\begin{equation}
		\gamma = -\left(2-\alpha - 2 \Delta \right), \hspace{0.5em}
		\gamma_{2} = -\left(2-\alpha - 2 \phi \right).
		\label{eq:scaling2}
	\end{equation}
	Eliminating $\Delta$ and $\phi$ in (\ref{eq:scaling1}) and (\ref{eq:scaling2}), we obtain the scaling relations:
	\begin{equation}
		\alpha+ 2\beta + \gamma =2, \hspace{0.5em} \alpha +2\beta_{2}+\gamma_{2}=2,
		\label{eq:Rushbrook}
	\end{equation}
	which are known as the Rushbrooke's scaling relation.
	The critical behaviors of the order parameter and susceptibilities along the various paths approaching the TCP in the three-dimensional space ($t,g,h$) can be derived from the scaling forms, (\ref{eq:scalingOP}) and (\ref{eq:scalingS}).
	In particular, considering a path such that $\abs{t}\to 0$ with $g=0$ for $\sigma$ and $\abs{t}\to 0$ with $h=0$ for $\sigma_{2}$, one finds that the order parameter and density are given by 
	\begin{equation}
		\sigma(h) = \abs{h}^{\beta/\Delta} \equiv \abs{h}^{1/\delta}, \hspace{0.5em} \sigma_{2}(g) = \abs{g}^{\beta_{2}/\phi} \equiv \abs{g}^{1/\delta_{2}}.
		\label{eq:delta}
	\end{equation}
	Here, we have used the fact that the order parameter must be regular on the path, such as ${\cal{M}}^{(\pm)}(0,x) \to x^{\beta/\Delta}$ for $x\to \infty$. 
	Combining with (\ref{eq:scaling1}), (\ref{eq:scaling2}), and (\ref{eq:delta}), we obtain the following scaling relations:
	\begin{equation}
		\gamma = \beta\left(\delta -1 \right), \hspace{0.5em} \gamma_{2} = \beta_{2} \left(\delta_{2}-1 \right),
		\label{eq:Widom}
	\end{equation}
	which are known as the Widom's scaling relation.
	Moreover, eliminating $\gamma$ and $\gamma_{2}$ in (\ref{eq:Rushbrook}) and (\ref{eq:Widom}), we obtain 
	\begin{equation}
		\alpha + \beta (\delta+1) =2, \hspace{0.5em} \alpha + \beta_{2} (\delta_{2}+1) =2,
		\label{eq:Griffiths}
	\end{equation}
	which are known as the Griffiths' scaling relation.
	
	The derivation of the scaling forms for $(\alpha_{t},\beta_{t},\delta,\gamma_{t})$ and $(\alpha_{t},\beta_{2t},\delta_{2},\gamma_{2t})$ is straightforward because the above discussion can also be applied to a path approaching the TCP at a finite angle with the phase boundary.
	
	In summary, the tricritical exponents $(\alpha,\phi,\Delta)$ characterize the critical phenomena at the TCP. 
	On the other hand, we define the exponents $(\alpha,\beta,\delta,\gamma)$ associated with the singular behaviors of physical quantities.
	These four exponents are not independent of each other and are related through the scaling relations because only the three exponents $(\alpha,\phi,\Delta)$ are independent.
	In other words, the scaling relations will serve as the consistency check for the validity of the scaling hypothesis.
	
	The scaling hypothesis can be applied to the correlation function which depends on the spatial coordinate $\vec{x}$.
	In momentum space, we assume that the correlation function is written in the following scaling form:
	\begin{equation}
		G(t,g;\vec{k}) = k^{\eta -2} {\cal{C}}^{(\pm)} \left(\frac{\abs{t}}{k^{1/\nu}},\frac{\abs{g}}{k^{1/\nu_{t}}} \right),
		\label{eq:scalingG}
	\end{equation}
	where $k=|\vec{k}|$, and ${\cal{C}}^{(\pm)}(x,y)$ is the scaling function.
	The exponent $\eta$ is called the anomalous dimension. At the TCP ($t=g=0$), this scaling form is given by
	\begin{equation}
		G(k)\propto k^{\eta-2}.
		\label{eq:etadef}
	\end{equation}
	Here, $\nu$ is defined by the critical behaviors of the correlation length, as we will explain later.
	If we first take the limit of $k\to 0$ in (\ref{eq:scalingG}), the correlation function corresponds to the susceptibility. Assuming that the scaling function is regular for $k\to 0$, we obtain the following scaling relations:
	\begin{equation}
		\gamma= \nu (2-\eta), 
		\label{eq:Fisher}
	\end{equation}
	which is known as the Fisher's scaling relation.
	Note that the similar discussion can be applied in the derivation of the scaling relations for $(\gamma_{2}, \nu_{2}, \eta_{2})$. 
	Here, one expects $\nu=\nu_{2}$ because the singular behavior of the correlation function is characterized by the single length scale, the correlation length.
	For a path approaching at a finite angle with the phase boundary, we obtain the Fisher's scaling relation of the exponents with the subscript $t$.
	
	\section{Landau theory of tricritical points} \label{sec:Landau}
	In this section, we briefly review tricritical behaviors in the framework of the Landau theory.
	In section \ref{subsec:crit1}, we begin with the free energy functional, the so-called Landau free energy, and derive the critical behaviors and the scaling relations.
	Here, we assume that the order parameter is spatially uniform and show the critical exponents ($\alpha,\beta,\delta,\gamma$).
	In section \ref{subsec:crit2}, we study the critical phenomena associated with the correlation function by considering the spatial dependence of the order parameter. 
	Here, we show the critical exponents ($\nu, \eta$).

	\subsection{Critical exponents from free energy} \label{subsec:crit1}
	In the vicinity of the TCP, we expand the free energy functional in powers of the order parameter up to a sixth-order term
	\begin{equation}
		\Phi(\sigma; T,\mu,m) = \Phi_{0} - m \sigma + \frac{a}{2}\sigma^{2} + \frac{b}{4}\sigma^{4}+ \frac{c}{6} \sigma^{6}, 
		\label{eq:free}
	\end{equation}
	where we assume that the first term $\Phi_{0}$ contributes only to a regular part of the free energy. 
	Furthermore, we also assume that $c$ is a positive constant, while $a$ and $b$ depend on $T$ and $\mu$. For our present purpose, we do not consider the $\sigma^{3}$ and $\sigma^{5}$ terms without loss of generality.
	The free energy is invariant under the transformations of $m\to-m$ and $\sigma\to-\sigma$, implying the phase diagram is symmetric with respect to the plane of $m=0$.
	Therefore, we refer to the plane of $m=0$ as the symmetric plane.
	The expectation value of the order parameter is given by the condition for a local minimum of the energy functional:
	\begin{equation}
		\frac{\partial \Phi}{\partial \sigma}=0 \hspace{0.5em} \Longleftrightarrow \hspace{0.5em} m=a\sigma +b\sigma^{3} + c\sigma^{5}
		\label{eq:expsigma}
	\end{equation}
	
	First, we consider the symmetric plane ($m=0$). In this case, the only solution is $\sigma=0$ for positive $a$ and $b$, which corresponds to the symmetry restored phase. For $a<0$ and $b>0$, on the other hand, $\sigma=0$ is the local maximum of $\Phi$, and we obtain the two local minimum solutions:
	\begin{equation}
		\sigma_{\pm} = \pm\left( \frac{-b + \sqrt{b^{2}-4ac}}{2c} \right)^{1/2},
		\label{eq:OP1}
	\end{equation}
	where the subscript $\pm$ denotes the sign of the right hand side. Since $\sigma\to 0$ in the limit of $a\to 0$, the line of $a=a_{\lambda}(b)=0$ can be identified as a line of critical points, the $\lambda$-line.
	For $b<0$, there are two specific values of $a$. One is $a=b^{2}/(4c)$, above which the non-zero solutions $\sigma_{\pm}$ appear, whereas below which $\sigma=0$ is the only solution. The other is the value at which the three phases coexist:
	\begin{equation}
		\Phi(\sigma_{\pm}) = \Phi(0) \hspace{0.5em} \Longleftrightarrow \hspace{0.5em} a=\frac{3b^{2}}{16 c}.
	\end{equation}
	In other words, the line of $a=a_{\tau}(b) = {3b^{2}}/{(16 c)}$ is the three-phase coexistence line, the triple line. The TCP is defined by the end point of the triple line, namely $a=b=0$.
	
	In the vicinity of the TCP, we assume that the coefficients $a$ and $b$ depend analytically on $T$ and $\mu$ and are expanded as
	\begin{eqnarray}
		a(T,\mu) &=& a_{1}(T-T_{t}) + a_{2}(\mu-\mu_{t}) +\cdots, \label{eq:a1}\\
		b(T,\mu) &=& b_{1}(T-T_{t}) + b_{2}(\mu-\mu_{t}) +\cdots, \label{eq:b1}
	\end{eqnarray}
	where $a_{i}$ and $b_{i}$ $(i=1,2)$ are constants. These constants need to satisfy some conditions depending on the geometry of the phase boundary. 
	Note that this crucial assumption leads to the behavior in the mean-field approximation.
	Furthermore, we rewrite the coefficients $a$ and $b$ with the scaling fields $t$ and $g$. Using the definition of the scaling fields $(\ref{eq:scalingfield})$, we can write
	\begin{eqnarray}
		a(t,g) &=& A g +\cdots, \\
		b(t,g) &=& B t + C g +\cdots,  
	\end{eqnarray}
	where $A>0$, $B>0$, and $C$ is a constant. Here, $a(t,g)$ does not contain a linear term in $t$ because the line of $g=0$ is the tangential line at the TCP and corresponds to the $\lambda$-line ($a=0$) in the linear level. 
	
	To study the critical behavior, various paths approaching the TCP are possible. 
	For example, let us consider a path $a\to -0$ ($T<T_{t}$) with $b=0$, which is not asymptotically tangential to the phase boundary. Then, we find that the critical behavior of the order parameter is given by
	\begin{equation}
		\sigma_{\pm} = \pm \left( \frac{\sqrt{-4ac}}{2c} \right)^{1/2} \propto \abs{g}^{1/4},
		\label{eq:beta1/4}
	\end{equation}
	where we obtain $\beta_{t}=1/4$.
	We consider another path along the triple line, $b\to-0$ with $a=a_{\tau}(b) = {3b^{2}}/{(16 c)}$.
	Then
	\begin{equation}
		\sigma_{\pm} = \pm\left( -\frac{b}{4c}\right)^{1/2} \propto \abs{t}^{1/2},
		\label{eq:beta1/2}
	\end{equation}
	where we obtain $\beta=1/2$. Note that the difference between these two values of $\beta_{t}$ and $\beta$ results from a choice of the path to the TCP as discussed in the previous section. 
	
	Here, we attempt to write the order parameter as the scaling form.
	For our convenience, we introduce new scaling fields $\tilde{g} \equiv a/ c^{1/3}$ and $\tilde{t}\equiv b / c^{2/3}$, ($\ref{eq:OP1}$) can be written as\footnote{The scaling fields $\tilde{t}$ and $\tilde{g}$, introduced instead of $t$ and $g$, are convenient to describe the scaling forms. However, when we consider the two typical paths, namely a path that is asymptotically tangential to the phase boundary and not tangential, the critical behaviors with either scaling field can be described in the same way. For instance, a path $a\to-0$ with $b=0$ corresponds to $g\to 0$ with $t\sim g$ and $\tilde{g}\to 0$ with $\tilde{t}=0$, respectively. Hence, we use $t$ and $g$ for the critical behaviors of each quantity, whereas $\tilde{t}$ and $\tilde{g}$ for the scaling forms.}
	\begin{equation}
		\tilde{\sigma}(\tilde{t},\tilde{g}) \equiv \sqrt{2} c^{1/6} \sigma_{\pm}  = \pm \left( \sqrt{\tilde{t}^{2}-4\tilde{g}}  - \tilde{t} \right)^{1/2},
		\label{eq:OP2}
	\end{equation}
	and this can be written as the scaling form
	\begin{equation}
		\tilde{\sigma}(\tilde{t},\tilde{g})= \pm\abs{\tilde{t}}^{\beta} {\cal{M}}^{(\pm)}\left( \frac{\tilde{g}}{\abs{\tilde{t}}^{\phi}} \right)=\pm \abs{\tilde{g}}^{\beta_{t}} {\cal{M}}^{(\pm)}_{t}\left( \frac{\tilde{t}}{\abs{\tilde{g}}^{\phi_{t}}} \right),
	\end{equation}
	where
	\begin{eqnarray}
		{\cal{M}}^{(\pm)}(x) &=& \left( \sqrt{1-4x} \mp 1 \right)^{1/2}, \\
		{\cal{M}}^{(\pm)}_{t}(y) &=& \left( \sqrt{y^{2}\mp4} - y \right)^{1/2},
	\end{eqnarray}
	with $\beta=1/2$, $\beta_{t}=1/4$, and $\phi= \phi_{t}^{-1}=2$.

	Now we study the critical behavior of the chiral susceptibility given by
	\begin{equation}
		\chi_{\rm ch} \equiv \left. \frac{\partial \sigma}{\partial m} \right|_{m=0}= \frac{1}{a+3b\sigma^{2}+5c\sigma^{4}},
	\end{equation}
	where we obtain this by differentiating (\ref{eq:expsigma}) with respect to $m$. 
	Here, $\sigma$ is the solution given by (\ref{eq:OP1}), and we omit the subscript $\pm$.
	Along the path considered in (\ref{eq:beta1/4}), the chiral susceptibility is given by
	\begin{equation}
		\chi_{\rm ch} = -\frac{1}{4a} \propto \abs{g}^{-1},
		\label{eq:chi1}
	\end{equation}
	where we find $\gamma_{t}=1$.
	In addition, along an asymptotically tangential path, such as along the triple line considered in (\ref{eq:beta1/2}), we find a different critical behavior,
	\begin{equation}
		\chi_{\rm ch} = -\frac{4 c}{b^{2}} \propto \abs{t}^{-2},
		\label{eq:chi2}
	\end{equation}
	where we find $\gamma=2$. The chiral susceptibility is rewritten with the new scaling fields as follows:
	\begin{equation}
		\tilde{\chi}_{\rm ch}(\tilde{t},\tilde{g}) \equiv c^{1/3}\chi_{\rm ch} = \frac{1}{\left( \tilde{t}^{2}-4\tilde{g} -\tilde{t}\sqrt{\tilde{t}^{2}-4\tilde{g}} \right)},
		\label{eq:chisus}
	\end{equation}
	and this can be written as
	\begin{equation}
		\tilde{\chi}_{\rm ch}(\tilde{t},\tilde{g})= \abs{\tilde{t}}^{-\gamma} {\cal{X}}^{(\pm)}\left( \frac{\tilde{g}}{\abs{\tilde{t}}^{\phi}} \right) =\abs{\tilde{g}}^{-\gamma_{t}} {\cal{X}}_{t}^{(\pm)}\left( \frac{\tilde{t}}{\abs{\tilde{g}}^{\phi_{t}}} \right),
	\end{equation}
	where 
	\begin{eqnarray}
		{\cal{X}}^{(\pm)}\left(x \right) &=& \left( 1-4x^{2}\mp \sqrt{1-4x^{2}} \right)^{-1},\\
		{\cal{X}}^{(\pm)}_{t}\left(y \right) &=& \left( y^{2}\mp 4- y \sqrt{y^{2}\mp4} \right)^{-1},
	\end{eqnarray}
	with $\gamma_{t}=\gamma /\phi = 1$.
	Note that the critical exponent $\gamma$ can be also defined in the symmetry restored phase ($\sigma=0$).
	In this case, $\chi_{\rm ch}= a^{-1}$ and the critical behavior is the same as in (\ref{eq:chi1}).

	Substituting the solution (\ref{eq:OP2}) to the free energy, we obtain the singular part of the free energy
	\begin{equation}
		\Phi(\tilde{t},\tilde{g}) = \frac{\tilde{t}(\tilde{t}^{2} - 6\tilde{g}) - (\tilde{t}^{2}-4\tilde{g})^{3/2} }{24},
	\end{equation}
	and the scaling form is given by
	\begin{equation}
		\Phi(\tilde{t},\tilde{g}) = \abs{\tilde{t}}^{2-\alpha} {\cal{G}}^{(\pm)} \left( \frac{\tilde{g}}{\abs{\tilde{t}}^{\phi}}\right) =\abs{\tilde{g}}^{2-\alpha_{t}} {\cal{G}}^{(\pm)}_{t} \left( \frac{\tilde{t}}{\abs{\tilde{g}}^{\phi_{t}}} \right),
		\label{eq:FEscaling}
	\end{equation}
	where
	\begin{eqnarray}
		{\cal{G}}^{(\pm)}(x)&=& \frac{\pm(1-6x)-(1-4x)^{3/2}}{24}, \\
		{\cal{G}}^{(\pm)}_{t}(y)&=& \frac{y(y^{2}\mp 6)-(y^{2} \mp 4)^{3/2}}{24},
	\end{eqnarray}
	with $\alpha=-1$ and $\alpha_{t}=1/2$.
	From the definition of $\alpha$ in (\ref{eq:FEscaling}), the critical exponents $\alpha$ and $\alpha_{t}$ are related to the singular behaviors of the second derivative of the free energy:
	\begin{equation}
		c_{g}\equiv -\frac{\partial^{2} \Phi}{\partial t^{2}} \propto
		\abs{t}^{-\alpha} \hspace{0.5em} (\text{along $\tau$-line}), \hspace{0.5em}
		\chi_{2}=-\frac{\partial^{2} \Phi}{\partial g^{2}} \propto
		\abs{g}^{-\alpha_{t}} \hspace{0.5em} (\text{along } t = 0).
		\label{eq:cv0}
	\end{equation}
	Here, $\chi_{2}$ is the susceptibility with respect to the density and $c_{g}$ is defined as the specific heat with the scaling field $g$ fixed.
	
	So far, we have discussed the critical behaviors in the symmetric plane ($m=0$). 
	For finite $m$, we can study another critical exponents from the behavior of the order parameter with respect to $m$.
	For $t=0$ and $g=0$, one finds the following relation from (\ref{eq:expsigma}) using the scaling field $h=m$,
	\begin{equation}
		\sigma_{\pm} \propto \abs{h}^{1/\delta},
		\label{eq:deltaL}
	\end{equation}
	with $\delta=5$.
	
	By differentiating the free energy with respect to the scaling field $g$, we obtain the density $\sigma_{2}$.
	Applying the similar procedure of the above discussion, we define the critical exponents associated with the density by
	\begin{equation}
		\sigma_{2} \propto 
		\left\{
		\begin{array}{ll}
			\abs{g}^{\beta_{2t}} & (\text{along } t = 0)\\
			\abs{t}^{\beta_{2}} & (\text{along $\tau$-line})
		\end{array}
		\right. , \hspace{0.5em}
		\chi_{2} \propto 
		\left\{
		\begin{array}{ll}
			\abs{g}^{-\gamma_{2t}} & (\text{along } t = 0)\\
			\abs{t}^{-\gamma_{2}} & (\text{along $\tau$-line})
		\end{array}
		\right.,
		\label{eq:beta2}
	\end{equation}
	where $\beta_{2t} = \beta_{2}/\phi = 1/\delta_{2}$ and $\gamma_{2t} = \alpha_{t}$ by definition (see (\ref{eq:delta}) and (\ref{eq:cv0})).
	In the conventional Landau theory, we obtain $\beta_{2}=1$, $\delta_{2}=2$, $\gamma_{2}=1$, and $\gamma_{2t}=1/2$.
	The scaling form for $\sigma_{2}$ is given by 
	\begin{equation}
		\sigma_{2}(\tilde{t},\tilde{g}) = \abs{\tilde{t}}^{\beta_{2}}{\cal{M}}^{(\pm)}_{2}\left( \frac{\tilde{g}}{\abs{\tilde{t}}^{\phi}}\right)=\abs{\tilde{g}}^{\beta_{2t}}{\cal{M}}^{(\pm)}_{2t}\left(\frac{\tilde{t}}{\abs{\tilde{g}}^{\phi_{t}}}\right),
	\end{equation}
	where
	\begin{eqnarray}
		{\cal{M}}^{(\pm)}_{2}(x) = \frac{1\pm\sqrt{1-4x}}{4}, \hspace{0.5em}
		{\cal{M}}^{(\pm)}_{2t}(y) = \frac{y-\sqrt{y^{2}\mp 4}}{4}.
	\end{eqnarray}
	Similarly, the scaling form for $\chi_{2}$ is given by
	\begin{equation}
		\chi_{2}(\tilde{t},\tilde{g}) = \abs{\tilde{t}}^{-\gamma_{2}}{\cal{X}}^{(\pm)}_{2}\left( \frac{\tilde{g}}{\abs{\tilde{t}}^{\phi}}\right)=\abs{\tilde{g}}^{-\gamma_{2t}}{\cal{X}}^{(\pm)}_{2t}\left( \frac{\tilde{t}}{\abs{\tilde{g}}^{\phi_{t}}}\right),
	\end{equation}
	where 
	\begin{eqnarray}
		{\cal{X}}^{(\pm)}_{2}(x) = -\frac{2}{\sqrt{1-4x}}, \hspace{0.5em}
		{\cal{X}}^{(\pm)}_{2t}(y) = -\frac{2}{\sqrt{y^{2}\mp 4}}.
	\end{eqnarray}

	In summary, we show the definitions and mean-field values of the critical exponents
	\begin{align}
		&    \sigma_{\pm} \propto \left\{
		\begin{array}{ll}
			\abs{g}^{\beta_{t}} & (\text{along } t = 0)\\
			\abs{t}^{\beta} & (\text{along $\tau$-line})
		\end{array}
		\right. , \hspace{0.5em}  \chi_{\rm ch} \propto 
		\left\{
		\begin{array}{ll}
			\abs{g}^{-\gamma_{t}} & (\text{along } t = 0)\\
			\abs{t}^{-\gamma} & (\text{along $\tau$-line})
		\end{array}
		\right. ,  \\
		& c_{g} \propto 
		\abs{t}^{-\alpha}  \hspace{0.5em} (\text{along $\tau$-line})
		, \hspace{1.5em} \sigma_{\pm} \propto \abs{h}^{1/\delta} \hspace{0.5em} (\text{at TCP}), \\
		& \sigma_{2} \propto\left\{
		\begin{array}{ll}
			\abs{g}^{1/\delta_{2}} & (\text{along } t = 0)\\
			\abs{t}^{\beta_{2}} & (\text{along $\tau$-line})
		\end{array}
		\right. , \hspace{0.5em} \chi_{2} \propto 
		\left\{
		\begin{array}{ll}
			\abs{g}^{-\alpha_{t}} & (\text{along } t = 0)\\
			\abs{t}^{-\gamma_{2}} & (\text{along $\tau$-line})
		\end{array}
		\right.,
	\end{align}
	with
	\begin{center}
		\begin{tabular}{ c | c | c | c | c   }
			$\alpha_{t}$ & $\beta_{t}$ & $\gamma_{t}$ &  $\delta$ & $\delta_{2}$  \\ \hline 
			$\frac{1}{2}$ & $\frac{1}{4}$ & 1 & 5 & 2 
		\end{tabular} \hspace{0.5em} (along $t=0$),
		\vspace{0.5em}\\
		\begin{tabular}{ c | c | c | c | c  }
			$\alpha$ & $\beta$ & $\gamma$ & $\beta_{2}$ & $\gamma_{2}$ \\ \hline 
			-1 & $\frac{1}{2}$ & 2 &  1 & 1 
		\end{tabular} \hspace{0.5em} (along $\tau$-line).
	\end{center}
	Note that the mean-field values of the critical exponents satisfy the scaling relations such as (\ref{eq:Rushbrook}), (\ref{eq:Widom}) and (\ref{eq:Griffiths}).

	\subsection{Critical exponents from correlation functions} \label{subsec:crit2}
	In the above discussion, we have neglected fluctuations in microscopic variables. To study the behavior of the correlation function, we consider the spatial dependence of the order parameter.
	Here, the order parameter is given by the expectation value of the microscopic fluctuating variable, $\phi(x)$:
	\begin{equation}
		\sigma(x)=\expval{\phi(x)}.
	\end{equation}
	If we assume that the source $m$ also depends on the spatial coordinate, the free energy functional is given by
	\begin{equation}
		\Phi = \int d^{3}x \left( -m(x)\phi(x) + \frac{1}{2}\left[ \nabla \phi(x) \right]^{2} + V\left[\phi(x) \right] \right),
	\end{equation}
	where 
	\begin{equation}
		V\left[\phi(x) \right]\equiv \frac{a}{2}\phi(x)^{2} + \frac{b}{4}\phi(x)^{4} + \frac{c}{6}\phi(x)^{6},
	\end{equation}
	and we ignore the constant term $\Phi_{0}$.
	If the source is spatially uniform, the local minimum of $\Phi$ gives a spatially uniform order parameter $\bar{\sigma}$ because the gradient term in $\Phi$ is non-negative.
	The equation of state is given by
	\begin{equation}
		m(x) = -\nabla^{2} \sigma(x) +V'\left[ \sigma(x)\right].
	\end{equation}
	Taking the derivative of this with respect to $m(y)$, we obtain 
	\begin{equation}
		\delta(x-y) = \left( -\nabla^{2} +V''\left(\sigma \right) \right) G(x-y),
	\end{equation}
	where
	\begin{equation}
		G(x-y) = \expval{\left(\phi(x)- \bar{\sigma}\right)\left(\phi(y)- \bar{\sigma}\right)} = \frac{\delta \sigma(x)}{\delta m(y)},
	\end{equation}
	which is the correlation function of the order parameter.
	Here, $m$ and $\sigma$ are taken to be spatially uniform after differentiation.
	The correlation function is given by the Ornstein-Zernicke form:
	\begin{equation}
		G(x) = \frac{1}{(2\pi)^{3}}\int d^{3}k \,e^{i k x} \frac{1}{k^{2}+\chi_{\rm ch}^{-1}},
	\end{equation}
	where $\chi_{\rm ch} = V''\left[\sigma \right]^{-1}$ is the chiral susceptibility.
	When $\abs{x}$ is large, the correlation function behaves as $G(x) \sim \exp\left( -\abs{x} / \sqrt{ \chi_{\rm ch} } \right)$, and the correlation length is given by $\xi = \sqrt{\chi_{\rm ch}}$.
	Thus, the divergence of the correlation length near the TCP is related to that of the chiral susceptibility:
	\begin{equation}
		\xi \propto \abs{t}^{-\nu} \sim \abs{t}^{-\gamma/2}, \hspace{0.5em} \xi \propto \abs{g}^{-\nu_{t}} \sim \abs{g}^{-\gamma_{t}/2},
		\label{eq:nu}
	\end{equation}
	with $\nu=1$ and $\nu_{t}=1/2$.
	In addition, the correlation function of the order parameter in Fourier space is given by
	\begin{equation}
		\tilde{G}(k) = \frac{1}{k^{2}+\chi_{\rm ch}^{-1}}.
	\end{equation}
	Since the chiral susceptibility diverges at the TCP, we find $ \tilde{G}(k)|_{\rm TCP}\sim k^{-2}$ and $\eta=0$.
	Furthermore, using (\ref{eq:chisus}), the correlation function with the scaling fields $\tilde{t}$ and $\tilde{g}$ is rewritten as
	\begin{equation}
		\tilde{G}(\tilde{t},\tilde{g},k) = \abs{k}^{\eta-2} {\cal C}^{(\pm)} \left(\frac{\abs{\tilde{t}}}{k^{1/\nu}}, \frac{\abs{\tilde{g}}}{k^{1/\nu_{t}}} \right),
		\label{eq:SHforG}
	\end{equation}
	where the scaling function is given by
	\begin{equation}
		{\cal C}^{(\pm)}(x,y) = \left(1+x^{2}-4y - x\sqrt{x^{2}-4y} \right)^{-1}.
	\end{equation}
	The correlation function of the density can also be written as the scaling hypothesis.
	The critical exponent $\eta_{2}$ for the correlation function of the density is different with $\eta$, whereas $\nu_{2}=\nu$ as discussed in the previous section.
	
	In summary, we show the definitions of the critical exponents associated with the correlation function and these mean-field values\,\cite{Riedel1972b,Bausch1972}:
	\begin{align}
		\xi \propto \left\{
		\begin{array}{ll}
			\abs{g}^{\nu_{t}} & (\text{along } t = 0)\\
			\abs{t}^{\nu} & (\text{along $\tau$-line})
		\end{array}
		\right. , \hspace{0.5em}  
		\tilde{G}(k) \propto k^{\eta-2}
		, \hspace{0.5em} \tilde{G}_{2}(k) \propto k^{\eta_{2}-2}, 
		\label{eq:nueta}
	\end{align}
	with
	\begin{center}
		\begin{tabular}{ c | c | c | c   }
			$\nu_{t}$ & $\nu$ & $\eta$ &  $\eta_{2}$  \\ \hline 
			$\frac{1}{2}$ & 1 & 0 & 1  
		\end{tabular}.
	\end{center}
	Here, $\tilde{G}_{2}$ is the correlation function of the density.
	The exponents $\eta$ and $\eta_{2}$ are evaluated at the TCP, ($t=g=0$).
	Note that the mean-field values of the critical exponents associated with the correlation functions satisfy the Fisher's scaling relation (\ref{eq:Fisher}).

	\subsection{Critical end point}
	For finite $m$, the order parameter at the critical end point has a finite value $\sigma_{0}$, and there is a two-phase coexistence line corresponding to the first-order phase transition line. At the critical end point ($T_{c}(m),\mu_{c}(m)$), the two minima and a maximum of the free energy functional coalesce. Hence, the critical end point is determined by the following conditions:
	\begin{eqnarray}
		&& \Phi' = -m + a_{m}\sigma_{0} +b_{m}\sigma_{0}^{3} + c \sigma_{0}^{5} =0, \\
		&& \Phi '' = a_{m} + 3 b_{m} \sigma_{0}^{2} +5 c \sigma_{0}^{4} =0, \\
		&& \Phi''' = 6 b_{m} \sigma_{0} + 20 c \sigma_{0}^{3} =0, 
	\end{eqnarray}
	where $a_{m}\equiv a\left(T_{c}(m), \mu_{c}(m) \right)$ and $b_{m}\equiv b\left( T_{c}(m), \mu_{c}(m) \right)$. Then, we obtain
	\begin{equation}
		\sigma_{0} = \sqrt{-\frac{3b_{m}}{10 c}}, \hspace{0.5em} a_{m}= \frac{9b_{m}^{2}}{20 c}, \hspace{0.5em} b_{m}=-\frac{5}{54^{1/5}}c^{3/5}m^{2/5}.
	\end{equation}
	Combining them with $(\ref{eq:a1})$ and $(\ref{eq:b1})$, the critical end point for small $m$ is written as
	\begin{eqnarray}
		T_{c}(m) - T_{t} &=&-\frac{45 a_{2} c^{1/5}}{4\left( 54\right)^{2/5}\left( b_{1}a_{2}-a_{1}b_{2}\right)} m^{2/5} + O\left( m^{4/5}\right), \label{eq:CEP1}\\
		\mu_{c}(m) - \mu_{t} &=&\frac{5 a_{1} c^{3/5}}{\left( 54\right)^{1/5}\left( b_{1}a_{2}-a_{1}b_{2}\right)} m^{2/5} + O\left( m^{4/5}\right),
		\label{eq:CEP2}
	\end{eqnarray}
	which gives the critical behavior of the line of the critical end points near the TCP.

	\section{Tricritical phenomena in holography} \label{sec:D3D7}
	In this section, we discuss the tricritical phenomena in the QCD-like holographic model, the D3/D7 model.
	The D3/D7 model presents ${\cal{N}} = 4$ supersymmetric Yang-Mills theory in $(3+1)$ dimensional spacetime with ${\cal{N}} =2$ hypermultiplets \cite{Karch:2002sh}. 
	For our purpose, we consider the system with a finite baryon number density and a finite magnetic field so that the TCP associated with the chiral phase transition emerges.
	
	\subsection{D3/D7 model}
	We consider a stack of $N_c$ D3-branes and a single D7-brane in the $(9+1)$ dimensional spacetime. The intersection of these branes is given by
	\begin{equation}
		\begin{array}{ccccccccccc}
			& 0 & 1 & 2 & 3 & 4 & 5 & 6 & 7 & 8 & 9 \\
			\mbox{D3} & \checkmark & \checkmark & \checkmark & \checkmark & - & - & - & - & - & - \\
			\mbox{D7} & \checkmark & \checkmark & \checkmark & \checkmark & \checkmark & \checkmark & \checkmark & \checkmark & - & -
		\end{array}
		\label{eq:intersection}
	\end{equation}
	Taking the large-$N_{c}$ limit with the large 't~Hooft coupling, the D3-branes can be considered as the source of supergravity fields in AdS$_{5}\times S^{5}$ in the gravity picture. 
	A finite-temperature system in dual field theory corresponds to the supergravity in the Schwarzschild-AdS$_{5} \times S^{5}$:
	\begin{equation}
		\dd s^{2} = \frac{1}{u^{2}}\left( -f(u) \dd t^{2} + \frac{\dd u^{2}}{f(u)} + \dd \vec{x}^{2} \right) + \dd \Omega_{5}^{2},
		\label{eq:BGmetric}
	\end{equation}
	where 
	\begin{equation}
		f(u) = 1 - \frac{u^{4}}{u_{\rm H}^{4}}.
	\end{equation}
	Here, $(t, \vec{x}) = ( t, x,y,z )$ denote the coordinates of the dual field theory in the (3+1) dimensional spacetime and $u$ denotes the radial coordinate of the AdS$_{5}$ geometry.
	For simplicity, the radius of AdS$_{5}$ and $S^{5}$ are set to be unity.
	The location of the black hole horizon is given by $u=u_{\rm H}$.
	$\dd \Omega_{5}^{2}$ is the line element of $S^{5}$ part, given by
	\begin{equation}
		\dd \Omega_{5}^{2} =  \dd\theta^{2} + \sin^{2}{\theta} \dd \psi^{2} + \cos^{2}\theta \dd \Omega_{3}^{2},
	\end{equation}
	where $\dd \Omega_{3}^{2}$ denotes the line element of the $S^{3}$ part. The D7-brane fills the AdS$_{5}$ part and wraps the $S^{3}$ part of $S^{5}$ as shown in (\ref{eq:intersection}). 
	The configuration of the D7-brane is determined by the embedding functions ($\theta,\psi$).
	
	The dynamics of the D7-brane is given by the Dirac-Born-Infeld (DBI) action:
	\begin{equation}
		S_{\rm D7} = - T_{\rm D7} \int \dd^{8} \xi  \sqrt{-\det \left(g_{ab}  + 2 \pi \alpha' F_{ab}\right)},
	\end{equation}
	where the D7-brane tension is given by $T_{\rm D7}= (2\pi)^{-7} (\alpha')^{-4} g_{\rm s}^{-1}$ with the string coupling constant $g_{\rm s}$ and the string length $l_{\rm s}= \alpha'^{1/2}$. 
	The induced metric on the D7-brane is given by
	\begin{equation}
		g_{ab} = \frac{\partial X^{M}}{\partial \xi^{a} } \frac{\partial X^{N}}{\partial \xi^{b} } G_{MN},
	\end{equation}
	where $X^{M}$ ($M,N=0,\cdots ,9$) denotes the target space coordinate and $\xi^{a}$ ($a,b=0,\cdots ,7$) denotes the worldvolume coordinate on the D7-brane. $G_{MN}$ denotes the background metric given by (\ref{eq:BGmetric}). The $U(1)$ gauge field strength on the D7-brane is defined by $2\pi \alpha' F_{ab}\equiv \partial_{a} A_{b} - \partial_{b} A_{a}$.
	For our purpose, we assume the following ansatz for fields:
	\begin{equation}
		\theta= \theta(u), \hspace{1em} \psi=0, \hspace{1em} A_{t}= a_{t}(u), \hspace{1em} A_{y}= B x,
	\end{equation}
	and the other components of the gauge fields are assumed to be zero. Then, the DBI action can be explicitly written as
	\begin{equation}
		S_{\rm D7}= -{\cal{N}} {\rm Vol_{4}} \int \dd u \cos^{3}\theta \sqrt{-g_{xx}\left( g_{xx}^{2} +B^{2}\right) \left(a_{t}'^{2} + g_{tt}g_{uu} \right) },
		\label{eq:S_D7}
	\end{equation}
	where
	\begin{equation}
		{\cal{N}} \equiv T_{\rm D7}(2 \pi^{2}) = \frac{N_{c} \lambda}{(2\pi)^{4}}.
	\end{equation}
	Here, we used the relation $4\pi g_{\rm s}  N_{c} \alpha'^{2} = \lambda \alpha'^{2} =  1$, where $\lambda = 4\pi g_{\rm s} N_{c} = g_{\rm YM}^{2} N_{c} $ is the 't~Hooft coupling, and ${\rm Vol_{4}} $ is the volume of the four dimensional spacetime ($t,\vec{x}$).
	
	Near the AdS boundary, the fields are asymptotically given by
	\begin{eqnarray}
		\sin\theta(u) &=& m u + c u^{3} + \cdots,  \\
		a_{t}(u) &=& \mu -\frac{\rho}{2}u^{2}  + \cdots, \label{eq:atAsym}
	\end{eqnarray}
	where, according to the AdS/CFT dictionary, $m$ and $c$ are related to the quark mass $m_{q}$ and chiral condensate $\expval{\bar{q}q}$, and $\mu$ and $\rho$ are related to the chemical potential $\mu_{q}$ and the expectation value of the baryon charge density in the dual field theory\,\cite{Mateos:2006nu,Mateos:2007vn,Kobayashi:2006sb}:
	\begin{equation}
		m_{q} = \frac{\lambda^{1/2}}{2\pi}m, \hspace{1em} \expval{\bar{q}q} = \frac{2N_{c}\lambda^{1/2}}{(2\pi)^{3}}c,
	\end{equation}
	and
	\begin{equation}
		\mu_{q} = \frac{\lambda^{1/2}}{2\pi}\mu, \hspace{1em} n_{q} = \frac{N_{c}\lambda^{1/2}}{(2\pi)^{3}}\rho. 
	\end{equation}
	Note that $\rho$ is related to the baryon number density which is given by $n_{b} = n_{q}/N_{c}$, where $n_{q}$ is the expectation value of the quark number density\,\cite{Kobayashi:2006sb}.
	Henceforth, we refer to $m$, $c$, $\mu$, and $\rho$ as the quark mass, the chiral condensate, the chemical potential, and the density, respectively.
	
	Since the DBI action contains only the derivative terms for the gauge fields, we obtain the conserved quantity from the equations of motion for $a_{t}$
	\begin{equation}
		\frac{\delta S_{D7} }{\delta F_{ut}} = \frac{ g_{xx}\left(g_{xx}^{2} +B^{2}\right)a_{t}' \cos^{3}\theta  }{\sqrt{-g_{xx}\left( g_{xx}^{2} +B^{2}\right) \left(a_{t}'^{2} + g_{tt}g_{uu} \right)}}.
	\end{equation}
	Substituting the asymptotic form (\ref{eq:atAsym}) near the AdS boundary, we find that this conserved quantity coincides with $\rho$ as $\rho = -\lim_{u\to 0} \delta S_{D7} / \delta F_{ut}$. 
	Thus, the solution is given by
	\begin{equation}
		a_{t}'(u) = \sqrt{\frac{-g_{tt}g_{uu} \rho^{2} }{\rho^{2}  +g_{xx}\left(g_{xx}^{2}+B^{2} \right)\cos^{6}\theta}}.
		\label{eq:atsol}
	\end{equation}
	If $\rho=0$, the solution is trivial and we obtain the constant $a_{t}$. 
	For a finite $\rho$, the chemical potential is given by
	\begin{equation}
		\mu\equiv \lim_{u\to 0} a_{t}(u) = \int^{u_{\rm H}}_{0}du \sqrt{\frac{-g_{tt}g_{uu} \rho^{2} }{\rho^{2}  +g_{xx}\left(g_{xx}^{2}+B^{2} \right)\cos^{6}\theta}}.
	\end{equation}
	
	For convenience, we perform the Legendre transformation of the action with respect to $a_{t}(u)$:
	\begin{equation}
		\tilde{S}_{D7} = S_{D7} - \int \dd^{8} \xi F_{ut} \frac{\delta S_{D7}}{\delta F_{ut}} = -{\cal{N}}{\rm Vol}_{4} \int \dd u\sqrt{-g_{tt}g_{uu}}\sqrt{\rho^{2}+g_{xx}(B^{2}+g_{xx}^{2})\cos^{6}\theta}.
		\label{eq:S_D7L}
	\end{equation}
	Then, we consider the equation of motion only for the embedding function $\theta(u)$.
	To determine the configuration of the D7-brane, we numerically solve the equation of motion for $\theta(u)$ obtained from (\ref{eq:S_D7L}). There are three types of solutions depending on the boundary condition in the bulk region: the {\it Minkowski embedding}, {\it black hole embedding}, and {\it critical embedding}\,\cite{Mateos:2006nu,Mateos:2007vn}.
	In the Minkowski embedding, the D7-brane does not reach the black hole horizon.
	In the black hole embedding, on the other hand, the D7-brane falls through the black hole horizon.
	The critical embedding is between the above two embeddings and forms a conical singularity at a point of the horizon.
	In the dual field theory picture, the Minkowski and black hole embedding are identified as the bound state of mesons, i.e., the quark-antiquark bound state, and the meson melting state, respectively\,\cite{Mateos:2006nu}.
	
	Since the system is invariant under a scale transformation, the three scale-free parameters $(\tilde{m},\tilde{T},\tilde{\mu})\equiv (m/\sqrt{B},\pi T/ \sqrt{2B}, \mu/\sqrt{B})$ characterize the system.
	To determine the thermodynamic stability of the solutions, we define the thermodynamic potential density with the on-shell action by
	\begin{equation}
		\Omega(\tilde{m},\tilde{T},\tilde{\mu}) \equiv \frac{-S_{\rm D7}}{{\cal N}{\rm Vol_{4}}} = \int_{0}^{u_{\rm H}} \dd u \sqrt{-g_{tt}g_{uu}g_{xx}^{2}} \frac{\left( g_{xx}^{2} +B^{2} \right) \cos^{6}\theta}{\sqrt{\left( g_{xx}^{2} +B^{2} \right) g_{xx}\cos^{6}\theta+ \rho^{2}}},
	\end{equation}
	where we rewrite the action as a function of $\rho$ by substituting (\ref{eq:atsol}) into (\ref{eq:S_D7}).
	Since the integration is divergent at the boundary ($u=0$), we introduce a UV cutoff at $u=\varepsilon$ with a small constant $\varepsilon$, and the divergence must be canceled by the counterterm.
	The counterterm in our coordinates\,\cite{Karch:2005ms,Karch:2007pd} is given by\footnote{Since the UV cutoff $\varepsilon$ is a dimensionful constant, it is natural to describe the logarithmic counterterm as $-(B^{2}/2 ) \log \kappa \varepsilon$, where $\kappa$ is a dimensionful constant such that $\kappa\varepsilon$ becomes dimensionless. The choice of $\kappa$ is scheme-dependent and provides a finite shift of the on-shell action. For simplicity, we choose $\kappa=1$ as in\,\cite{Jensen:2010vd}. Other choices are discussed in \cite{Matsumoto:2018ukk,Imaizumi:2019byu}.}
	\begin{eqnarray}
		L_{\rm ct} = L_{1} +L_{2} +L_{f} +L_{F},
	\end{eqnarray}
	where
	\begin{align}
		&L_{1}=\frac{1}{4}{\cal{N}}\sqrt{-\det\gamma_{ij}}, \hspace{0.5em} L_{2} = -\frac{1}{2}{\cal{N}}\sqrt{-\det\gamma_{ij}} \theta(\varepsilon)^{2}, \\
		&L_{f} = \frac{5}{12}{\cal{N}}\sqrt{-\det\gamma_{ij}}\theta(\varepsilon)^{4}, \hspace{0.5em} L_{F}=-\frac{1}{4}{\cal{N}}(2\pi \alpha')^{2}\sqrt{-\det\gamma_{ij}}F^{2}\log \varepsilon.
	\end{align}
	Here, $\gamma_{ij}$ is the induced metric on the $u=\varepsilon$ slice and $(2\pi\alpha')^{2}F^{2} = 2 B^{2}$ in our setup.
	By adding the counterterm to the on-shell action, the divergence at the boundary is renormalized.
	Thus, we can determine the thermodynamic stability and the phase transition points for each parameter $(\tilde{m},\tilde{T},\tilde{\mu})$.
	
	\subsection{Phase diagram}
\begin{figure}[tbp]
		\centering 
		\includegraphics[width=10cm]{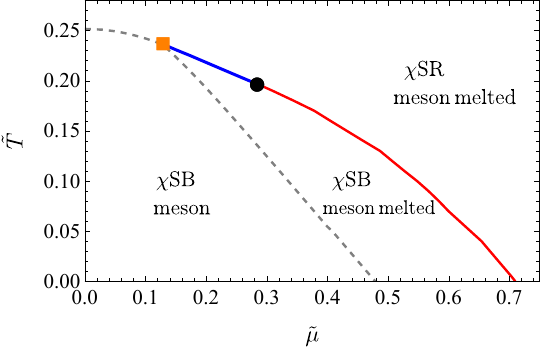}
		\caption{The phase diagram for $\tilde{m}=0$. The black filled circle and orange filled square denote the TCP associated with the chiral phase transition and the critical point of the transition between the meson bound state and melted state, respectively. The blue and red solid lines denote the first- and second-order phase transition lines, respectively. The gray dashed line denotes the second-order phase transition line between the meson bound state and melted state, but we do not focus on this transition in this paper.}
		\label{fig:phase0}
	\end{figure}

	For the purpose for studying the chiral symmetry breaking, we first consider a massless quark $\tilde{m}=0$. 
	For $\tilde{m}=0$, there are two different types of solution in terms of the configuration of the D7-brane: the flat configuration ($\theta=0$) and the bending configuration ($\theta\neq 0$). 
	The solutions can be distinguished by the $SO(2)$ rotational symmetry, where $\psi$ denotes the rotation angle. 
	The $SO(2)$ rotational symmetry in the bulk theory corresponds to the $U(1)$ chiral symmetry in the dual field theory \cite{Erdmenger:2007bn}. 
	Thus, these two configurations are identified as the chiral symmetry restored ($\chi{\rm SR}$) phase and broken ($\chi{\rm SB}$) phase, respectively.
	In fact, the chiral condensate $c$ obviously vanishes for $\theta=0$, and becomes finite for $\theta\neq 0$.
	Note that the flat configuration is always the black hole embedding, although the bending configuration is either the Minkowski or black hole embedding depending on the parameters.
	Figure \ref{fig:phase0} shows the phase diagram for $\tilde{m}=0$.
	As shown, there are three different phases: the meson melted $\chi{\rm SR}$, meson melted $\chi{\rm SB}$, and meson $\chi{\rm SB}$ phase.
	In this study, we focus only on the meson melted $\chi{\rm SR}$ and the meson melted $\chi{\rm SB}$ phase because we are interested in the critical phenomena at the TCP associated with the chiral symmetry breaking (black filled circle in Figure \ref{fig:phase0}).
	The blue and red solid lines denote the first- and second-order phase transition lines, respectively\footnote{Compared to the QCD phase diagram in the chiral limit, the transitions' orders are reversed in the D3/D7 model as discussed in \cite{Evans:2010iy}. At zero temperature, for example, the transition with chemical potential is second-order in our model, whereas first-order in QCD \cite{Halasz:1998qr}. However, the critical phenomena at the TCP should be independent of the explicit structure of the phase diagram.}.
	
\begin{figure}[tbp]
		\centering 
		\includegraphics[width=10cm]{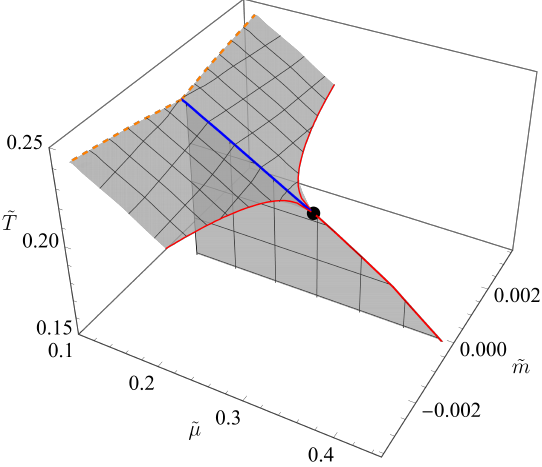}
		\caption{The three-dimensional phase diagram. The black dot denotes the TCP associated with the chiral transition. The three red solid lines denote the second-order phase transition lines: the $\lambda$-line for $\tilde{m}=0$ and two lines of the critical endpoints for $\tilde{m}>0$ and $\tilde{m}<0$. The blue solid line for $\tilde{m}=0$ denotes the first-order phase transition line, called the triple line. The two orange dashed lines for $\tilde{m}\neq 0$ denote the lines of another critical point associated with the transition between the meson bound state and melted state. The three gray surfaces denote the set of the first-order phase transition points.}
		\label{fig:phase3d}
	\end{figure}

	For $\tilde{m}\neq 0$, the chiral condensate is always finite because the chiral symmetry is explicitly broken by the quark mass. 
	That is why the $\lambda$-line for $\tilde{m}=0$ is replaced by the crossover, and the first-order phase transition line and the critical point are appeared for finite $\tilde{m}$.
	The set of critical points for $m\neq 0$ forms the critical line, and ends at the TCP for $\tilde{m}\to 0$.
	In this way, the three second-order phase transition lines, the $\lambda$-line for $\tilde{m}=0$ and two lines of critical end points for $\tilde{m}>0$ and $\tilde{m}<0$, end at the TCP.
	Figure \ref{fig:phase3d} shows the three-dimensional phase diagram near the TCP.

	\subsection{Tricritical phenomena from background}
	We study the critical phenomena near the TCP, and numerically determine the critical exponents introduced in the previous section.
	In analogy to (\ref{eq:scalingfield}), we employ the following scaling fields in our model:
	\begin{equation}
		t=\tilde{T}-\tilde{T}_{t}, \hspace{0.5em} g= \left(\tilde{\mu}-\tilde{\mu}_{t} \right) + A_{0}t, \hspace{0.5em} h=\tilde{m}.
	\end{equation}
	
\begin{figure}[tbp]
		\centering 
		\includegraphics[width=10cm]{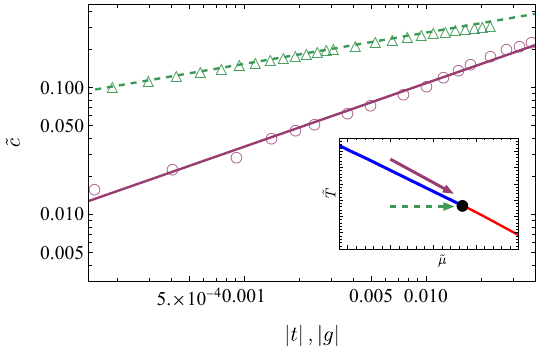}
		\caption{The critical behaviors of the chiral condensate for the two different paths to the TCP in log scale. Inset shows the phase diagram for $h=0$ and the two arrows denote the two different paths. One path is along the triple line, which is asymptotically tangential to the phase boundary (purple solid arrow). The other is the path with $t=0$ (green dashed arrow). The open circles and triangles denote the numerical plots for each path, respectively, and the purple solid and green dashed lines show the fitting results.}
		\label{fig:beta}
	\end{figure}

	First, we analyze the critical behavior of the order parameter, the chiral condensate, in the symmetric plane ($h=0$).
	Figure \ref{fig:beta} shows that the critical behaviors of the chiral condensate for two different paths to the TCP: the paths along the triple line and with $t=0$ fixed.
	We find that the critical behaviors of the chiral condensate are described as
	\begin{equation}
		\tilde{c} \propto \abs{t}^{\beta} , \hspace{1em} \tilde{c} \propto \abs{g}^{\beta_{t}},
	\end{equation}
	where $\tilde{c} \equiv c/B^{3/2}$. We obtain $\beta \approx 0.501$ and $\beta_{t} \approx 0.247$, which agree with the mean-field values, i.e. (\ref{eq:beta1/2}) and (\ref{eq:beta1/4}), respectively.
	
\begin{figure}[tbp]
		\centering 
		\includegraphics[width=10cm]{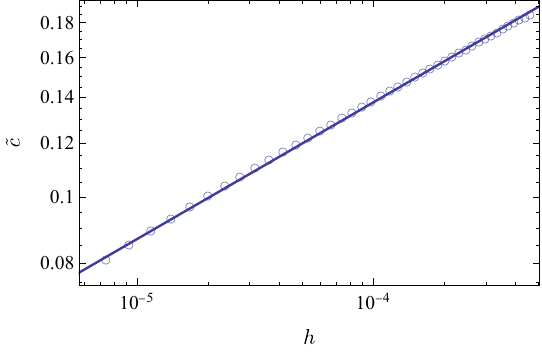}
		\caption{The critical behavior of the chiral condensate as a function of the quark mass in log scale. The open circles denote the numerical plots, and the solid line denotes the fitting result.}
		\label{fig:delta}
	\end{figure}

	Second, we analyze the critical behavior of the chiral condensate as a function of $h$ with $t=g=0$. 
	Figure \ref{fig:delta} shows the relation between $\tilde{c}$ and $h$ near the TCP in log scale.
	The critical behavior is given by
	\begin{equation}
		\tilde{c} \propto h^{1/\delta},
	\end{equation}
	and we obtain $\delta^{-1}\approx 0.200$ from the fitting of the numerical data. The value of the critical exponent also agrees with the mean-field value, $\delta = 5$, i.e., (\ref{eq:deltaL}).
	
\begin{figure}[tbp]
		\centering 
		\includegraphics[width=7cm]{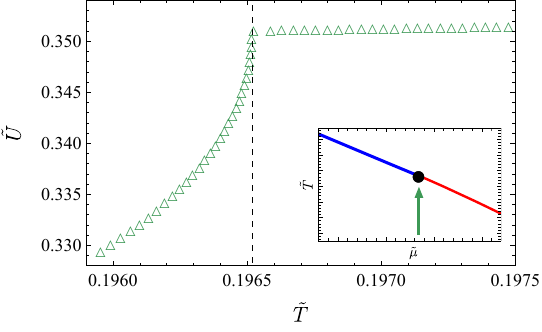}
		\includegraphics[width=7cm]{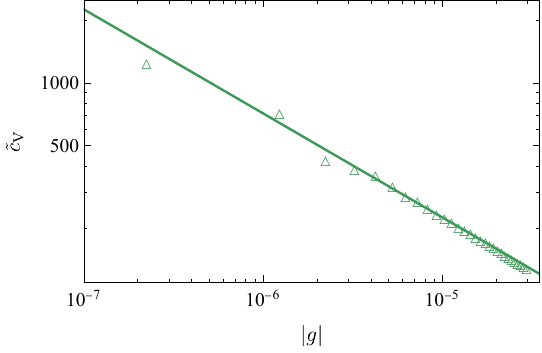}
		\caption{The left panel shows the critical behavior of the internal energy as a function of temperature. The inset shows the phase diagram, and the arrow denotes the path with the chemical potential fixed. The vertical dashed line denotes the tricritical temperature. The right panel shows the critical behavior of the specific heat in log scale. The open triangles and solid line denote the numerical data and the fitting result, respectively.}
		\label{fig:UTplot}
	\end{figure}

\begin{figure}[htbp]
		\centering 
		\includegraphics[width=10cm]{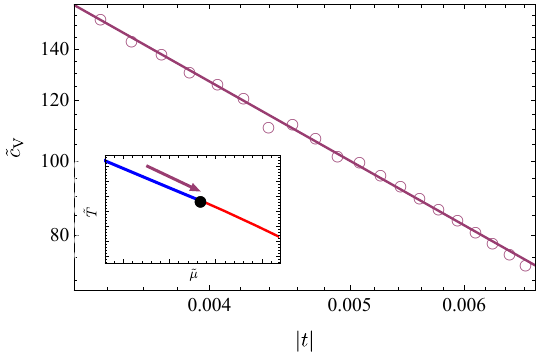}
		\caption{The plot shows the critical behavior of the specific heat along the triple line in log scale. 
			The open circles and the solid line denote the numerical data and the fitting result, respectively. The inset shows the phase diagram, and the arrow denotes the path along the triple line.
		}
		\label{fig:alphatau}
	\end{figure}

	Third, we analyze the critical behavior of the specific heat. Using the thermodynamic relations, the specific heat at constant chemical potential is given by
	\begin{equation}
		c_{\rm V} \equiv -T \frac{\partial^{2} \Omega}{\partial T^{2}} = T\frac{\partial S}{\partial T} = \frac{\partial U}{\partial T},
		\label{eq:cv}
	\end{equation}
	where $S=-\partial \Omega / \partial T$ is the entropy and $U = \Omega + \mu \rho + TS$ is the internal energy of the system.
	Thus, we can read off the specific heat from the slope of the internal energy with respect to temperature if we fix the chemical potential.
	The details of the calculation in the D3/D7 model are discussed in \cite{Mateos:2007vn,Kobayashi:2006sb}. 
	Figure \ref{fig:UTplot} shows the critical behavior of the internal energy (left panel) and the specific heat (right panel) with $\mu=\mu_{t}$.
	For the path with the chemical potential fixed, the specific heat is divergent at the TCP and we obtain
	\begin{equation}
		\tilde{c}_{\rm V} \propto \abs{g}^{-0.501},
	\end{equation}
	where $\tilde{c_{V}}\equiv c_{\rm V}/B^{3/2}$.
	On the other hand, Figure \ref{fig:alphatau} shows the critical behavior of the specific heat for the path along the triple line.
	We find that the specific heat along the triple line becomes divergent near the TCP with
	\begin{equation}
		\tilde{c}_{\rm V} \propto \abs{t}^{-1.063}.
	\end{equation}
	This value of the exponent seems not to be the mean-field value, (\ref{eq:cv0}).
	However, this is not inconsistent for the following reason.
	The specific heat at constant chemical potential should contain the contributions of three different susceptibilities: $c_{g} = \partial^{2} \Omega / \partial t^{2}$, $\partial^{2} \Omega / \partial t \partial g$, and $\chi_{2}=\partial^{2} \Omega / \partial g^{2}$ in terms of the scaling fields.
	Since the strongest singularity is that of $\chi_{2}$, the specific heat at constant chemical potential normally diverges at the TCP with the exponent $\gamma_{2}$ or $\gamma_{2t}$.
	For this reason, the critical behavior of the specific heat for each path represents the exponent $\gamma_{2t}=\alpha_{t}=1/2$ and $\gamma_{2}=1$, respectively.
	Note that we study the critical behavior of the specific heat for the path with $\mu=\mu_{t}$ instead of $t=0$ as in (\ref{eq:cv0}).
	We do not verify the value of $\alpha$ for the path along the triple line from the critical behavior of the specific heat because $c_{g}$ contributes the regular correction to $c_{\rm V}$.
	However, all other exponents so far take the mean-field values, and we verify that the Griffiths' scaling relation (\ref{eq:Griffiths}) are satisfied for $(\alpha_{t},\beta_{t},\delta)$.
	If we assume that the Griffiths' scaling relation is also valid for the path along the triple line, we conclude $\alpha=-1$ from ($\beta,\delta$)=(1/2, 5).
	
\begin{figure}[tbp]
		\centering 
		\includegraphics[width=7cm]{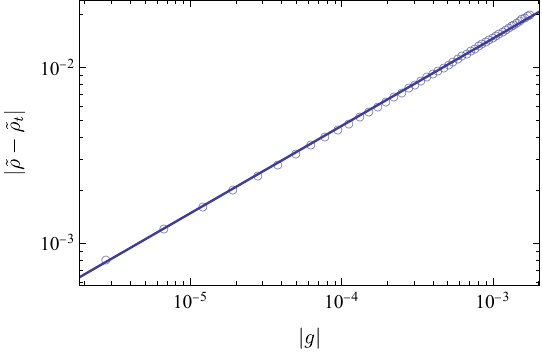}
		\includegraphics[width=6.5cm]{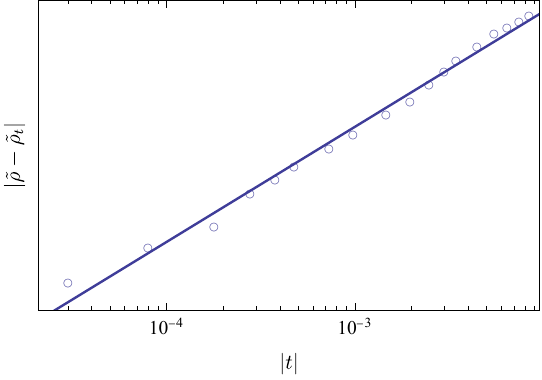}
		\caption{The plots show the critical behaviors of the density for a path with $t=0$ (left panel) and along the triple line (right panel). }
		\label{fig:beta2}
	\end{figure}

	Lastly, we study the critical exponents associated with the density.
	Since the density at the TCP is finite, we define the corresponding critical exponents by
	\begin{equation}
		\abs{\tilde{\rho}-\tilde{\rho}_{t}} \propto 
		\left\{
		\begin{array}{ll}
			\abs{g}^{1/\delta_{2}} & (\text{along } t = 0)\\
			\abs{t}^{\beta_{2}} & (\text{along $\tau$-line})
		\end{array}
		\right.,
	\end{equation}
	where $\tilde{\rho}=\rho/B^{3/2}$ and $\tilde{\rho}_{t}$ is the critical value of $\tilde{\rho}$ at the TCP.
	Figure \ref{fig:beta2} shows the critical behaviors of the density near the TCP.
	By fitting the numerical data, we obtain $\beta_{2}\approx1.021$ and $\delta_{2}^{-1}\approx0.498$, which agree with the mean-field values, i.e. (\ref{eq:beta2}).
	In addition, we verify that the Griffiths' scaling relation (\ref{eq:Griffiths}) is satisfied with $\alpha=-1$.
	
	So far, we have determined the critical exponents ($\alpha, \alpha_{t}, \beta, \beta_{2}, \beta_{t}, \delta, \delta_{2}$) at the TCP.
	Although we can compute the chiral susceptibility and the baryon number susceptibility by differentiating the order parameter and density with respect to the conjugate sources, and determine critical exponents ($\gamma, \gamma_{t}, \gamma_{2}$), we leave it in the next section because it can be also computed from the correlation functions.
	
	\subsection{Tricritical phenomena from correlation functions}
	In this section, we study the critical exponents associated with the correlation functions.
	In this paper, we consider the two-point correlation function of the chiral condensate $c$ and the density $\rho$.
	To compute the correlation functions, we consider small perturbations of the fields: $\theta \to \theta +\delta\theta$ and $a_{\mu}\to a_{\mu} +\delta a_{\mu}$.
	In general, we assume that perturbations can be written as the plane wave forms:
	\begin{equation}
		\delta\theta(t,u,\vec{x})=\vartheta(u) e^{-i \omega t + i k z}, \hspace{0.5em} \delta a_{\mu}(t,u,\vec{x})={\cal{A}}_{\mu}(u) e^{-i \omega t + i k z}
	\end{equation}
	with momentum $\vec{k}=(0,0,k)$, for simplicity.
	According to the holographic prescription\,\cite{Kovtun:2005ev}, we can compute the correlation functions, $\expval{c(x)c(y)}$ and $\expval{\rho(x)\rho(y)}$, by solving the equations of motion for the perturbations in bulk with the appropriate boundary conditions.
	The correlation function of the order parameter is characterized by the correlation length $\xi$ that diverges at the TCP.
	However, the correlation function of the density does not have a singular behavior if the density does not couple to the chiral condensate.
	Conversely, if the density and chiral condensate are coupled, the correlation function of the density is also characterized by the correlation length and expected to show the singular behavior.
	This is the case in the $\chi{\rm SB}$ phase as we will discuss later.
	
	\subsubsection{\texorpdfstring{$\chi{\rm SR}$}{TEXT} phase}
	In the $\chi{\rm SR}$ phase, the background solutions are explicitly written as
	\begin{equation}
		\theta(u)=0,\hspace{0.5em} a_{t}'(u)=\frac{\rho u }{\sqrt{\rho^{2}u^{6} +B^{2} u^{4} +1 }},
	\end{equation}
	where we have substituted the induced metric components into (\ref{eq:atsol}).
	Expanding the DBI action up to the second order in perturbations, we obtain
	\begin{align}
		S_{(2)}=\int du \frac{\sqrt{\zeta}}{u} \left[ \frac{f}{2u^{2}}\vartheta'^{2} \right. + & \frac{1}{u^{4}}\left( \frac{\omega^{2}u^{2}}{f} - \frac{(3+k^{2}u^{2})(1+B^{2}u^{4})}{\zeta} \right)\vartheta^{2}  \nonumber\\
		& \left. -\frac{\zeta}{1+B^{2}u^{4}} {\cal{A}}_{t}'^{2}+ f {\cal{A}}_{z}'^{2} + \frac{(k {\cal{A}}_{t} + \omega {\cal{A}}_{z})^{2}}{f} +\cdots \right],
	\end{align}
	where $\zeta = \rho^{2}u^{6} +B^{2} u^{4} +1$ and we choose the gauge $\delta a_{u}=0$. The ellipsis denotes the terms that contain ${\cal{A}}_{x}$ and ${\cal{A}}_{y}$.
	Here, one can see that the perturbation of $\theta$ is decoupled from the other perturbations, but the perturbation of $a_{t}$ is coupled to that of $a_{z}$.
	As mentioned above, in the $\chi {\rm SR}$ phase, the correlation function of the density does not show a singular behavior at the TCP because the perturbations of $\theta$ and $a_{t}$ are decoupled. 
	This is consistent with the fact that the susceptibility of the quark number, which corresponds to the correlation function of the density with $\omega=k=0$, is regular at the TCP in the $\chi$SR phase \cite{Hatta:2002sj}.
	Introducing the gauge-invariant field, $k {\cal{A}}_{t} +\omega {\cal{A}}_{z}$, it shows the so-called holographic zero sound mode at small temperature as discussed in\,\cite{Karch:2009zz,Goykhman:2012vy,Davison:2011ek}. 
	For this reason, we focus only on the perturbation of $\theta$ in the following.
	
	The equation of motion for $\vartheta$ is given by
	\begin{align}
		\vartheta''+\left(\frac{f'}{f}- \frac{3+B^{2}u^{4}}{u\zeta} \right)\vartheta'+ \left(\frac{\omega^{2}}{f^{2}} +\frac{(3-k^{2}u^{2})(1+B^{2}u^{4})}{u^{2}f\zeta}\right) \vartheta =0.
	\end{align}
	For our purposes, we impose the ingoing-wave boundary condition at the horizon:
	\begin{equation}
		\vartheta(u) \sim (u_{\rm H}-u)^{-i\frac{\omega}{4\pi T}} \vartheta_{\rm reg}(u),
	\end{equation}
	where $\vartheta_{\rm reg}$ is a regular part at the horizon $u=u_{\rm H}$.
	Near the boundary, $\vartheta$ can be written in the following asymptotic form:
	\begin{align}
		&\vartheta(u) = \vartheta^{(0)}u + \vartheta^{(1)} u^{3} + \cdots, 
	\end{align}
	where $\vartheta^{(0)}$ and $\vartheta^{(1)}$ are the non-normalizable mode and the normalizable mode, respectively. 
	Following the analysis of critical phenomena in \cite{Maeda:2009wv}, we assume that each mode can be expanded as a function of $(\omega,k)$,
	\begin{eqnarray}
		\vartheta^{(0)}&\sim& \vartheta^{(0)}_{0} + \omega \vartheta^{(0)}_{(1,0)}+ k^{2} \vartheta^{(0)}_{(0,1)}, \\
		\vartheta^{(1)}&\sim& \vartheta^{(1)}_{0} + \omega \vartheta^{(1)}_{(1,0)}+ k^{2} \vartheta^{(1)}_{(0,1)}.
	\end{eqnarray}
	Using these expressions, the retarded Green's function is given by
	\begin{eqnarray}
		G^{R}_{\vartheta \vartheta}(\omega, k) \sim \frac{\vartheta^{(1)}}{\vartheta^{(0)}} \sim
		\frac{\vartheta^{(1)}_{0} + \omega \vartheta^{(1)}_{(1,0)}+ k^{2} \vartheta^{(1)}_{(0,1)}}{\vartheta^{(0)}_{0} + \omega \vartheta^{(0)}_{(1,0)}+ k^{2} \vartheta^{(0)}_{(0,1)}}  
		\sim 
		\frac{\vartheta^{(1)}_{0}/\vartheta^{(0)}_{(0,1)}}{-iq\omega + k^{2} +1/\xi^{2}},
		\label{eq:Green}
	\end{eqnarray}
	where $q\equiv i \vartheta^{(0)}_{(1,0)}/\vartheta^{(0)}_{(0,1)}$, and $\xi\equiv \sqrt{\vartheta^{(0)}_{(0,1)}/\vartheta^{(0)}_{0}}$ is the correlation length. 
	
\begin{figure}[tbp]
		\centering 
		\includegraphics[width=7cm]{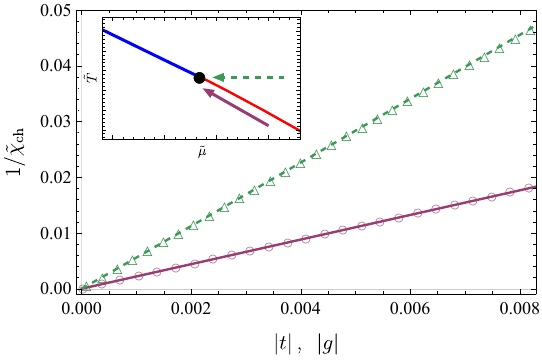}
		\includegraphics[width=7cm]{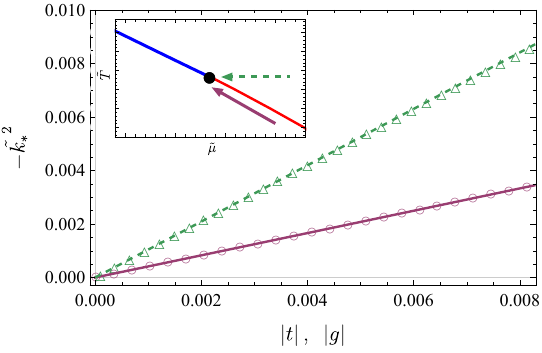}
		\caption{The critical behavior of $1/\tilde{\chi}_{\rm ch}$ (left panel) and $-\tilde{k}_{*}^{2}$ (right panel) along the $\lambda$-line (open circles and solid purple line) and the path with $t=0$ fixed (open triangles and dashed green line). The inset shows the phase diagram and two arrows denote the two different paths.}
		\label{fig:gnplot}
	\end{figure}

	If we take $\omega\to 0$, $G^{R}_{\vartheta \vartheta}(k) \propto 1/(k^{2}+\xi^{-2})$. Then, if we find the poles of the retarded Green's function, $k_{*}$, near the TCP, we can explore the critical behavior of the correlation length because $\xi^{2} \sim - k_{*}^{-2}$.
	Furthermore, if we take $\omega \to 0$ and $k\to 0$ in (\ref{eq:Green}), the retarded Green's function corresponds to the chiral susceptibility: $\chi_{\rm ch}=\vartheta_{0}^{(1)}/\vartheta_{0}^{(0)}$.
	Figure \ref{fig:gnplot} shows the critical behavior of $1/\tilde{\chi}_{\rm ch}$ and $-\tilde{k}_{*}^{2}$, where $\tilde{\chi}_{\rm ch}=\chi_{\rm ch}/B$ and $\tilde{k}_{*} = k_{*}/\sqrt{B}$, for the two different paths.
	These critical behaviors are related to the critical exponents $\gamma$ and $\nu$, respectively.
	Here, we consider the path along the $\lambda$-line as a tangential path.
	This is because the order parameter is zero along the $\lambda$-line ($\chi{\text{SR}}$), whereas the finite order parameter coexists along the triple line.
	By fitting the numerical data, we obtain
	\begin{equation}
		\tilde{\chi}_{\rm ch} \propto
		\left\{
		\begin{array}{ll}
			\abs{t}^{-\gamma_{\lambda}} & (\text{along $\lambda$-line})\\
			\abs{g}^{-\gamma_{t}} & (\text{along $t=0$})
		\end{array}
		\right. ,\hspace{0.5em} \tilde{\xi} \propto 
		\left\{
		\begin{array}{ll}
			\abs{t}^{-\nu_{\lambda}} & (\text{along $\lambda$-line})\\
			\abs{g}^{-\nu_{t}} & (\text{along $t=0$})
		\end{array}
		\right.,
		\label{eq:gammaR}
	\end{equation}
	with $\gamma_{\lambda}\approx 1.009$, $\gamma_{t}\approx1.001$, $\nu_{\lambda}\approx0.502$, and $\nu_{t}\approx0.502$. 
	Here, we introduce $\gamma_{\lambda}$ and $\nu_{\lambda}$ to distinguish them from the exponents along the triple line.
	These values take the mean-field values because the chiral susceptibility is written as $\tilde{\chi}_{\rm ch} \propto \abs{g}^{-1}$ in the $\chi$SR phase.
	
\begin{figure}[tbp]
		\centering 
		\includegraphics[width=7cm]{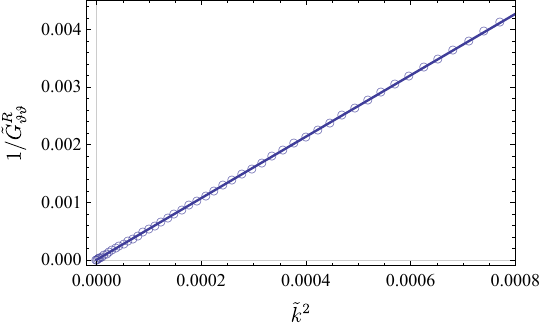}
		\includegraphics[width=7cm]{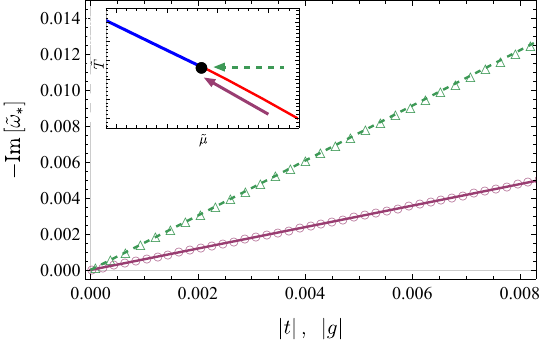}
		\caption{The left panel shows the critical behavior of $1/\tilde{G}_{\vartheta\vartheta}^{R}$ as a function of $\tilde{k}^{2}$ at the TCP. The open circles denote the numerical data, and the solid line denotes the fitting result. The right panel shows the critical behavior of $\tilde{\omega}_{*}$ along the $\lambda$-line (open circles and solid purple line) and the path with $t=0$ fixed (open triangles and dashed green line). The inset shows the phase diagram and two arrows denote the two different paths.}
		\label{fig:eta}
	\end{figure}

	Moreover, the retarded Green's function with $\omega\to 0$ behaves as $G^{R}_{\vartheta\vartheta}(k)\propto k^{-2}$ at the TCP because the correlation length is divergent there.
	From the definition of the critical exponent $\eta$ in (\ref{eq:etadef}), we assumed $\eta=0$ in the form of (\ref{eq:Green}).
	We numerically confirm $\eta=0$ from the critical behavior of the retarded Green's function as a function of $k$ at the TCP as shown in the left panel of Figure \ref{fig:eta}.
	Here, the tilde denotes the dimensionless variables, $\tilde{G}_{\vartheta\vartheta}^{R}=G_{\vartheta\vartheta}^{R}/B$.
	By fitting the numerical data, we obtain 
	\begin{equation}
		\tilde{G}^{R}_{\vartheta\vartheta}(k)\propto \tilde{k}^{\eta-2},
	\end{equation}
	with $\eta\approx 0.005$.
	Together with the critical exponents ($\gamma,\nu$), these values satisfy the Fisher's scaling relation (\ref{eq:Fisher}).
	
	In addition to the above critical exponents determined by the static correlation function, one can determine another critical exponent for finite $\omega$.
	If we take $k\to 0$ in (\ref{eq:Green}), we find $G^{R}_{\vartheta\vartheta}(\omega) \propto 1/(-iq\omega+\xi^{-2})$.
	Since the relaxation time is obtained from the imaginary part of the pole, $\tau_{k=0}^{-1}\sim - \Im\omega_{*} $, where $\omega_{*}$ is the pole of (\ref{eq:Green}) with $k=0$, we can determine the dynamic critical exponent $z$ defined by the critical behavior of the relaxation time: $\tau_{k=0}\propto\xi^{z}$.  
	We have also assumed that $z=2$ in (\ref{eq:Green}), and numerically confirm this in the right panel of Figure \ref{fig:eta}.
	We find 
	\begin{equation}
		\tau_{k=0}^{-1} \sim -\Im \omega_{*} \propto \xi^{-z} \sim 
		\left\{
		\begin{array}{ll}
			\abs{t}^{z\nu} & (\text{along $\lambda$-line})\\
			\abs{g}^{z\nu_{t}} & (\text{along $t=0$})
		\end{array}
		\right.,
		\label{eq:znu}
	\end{equation}
	where  we have assumed that the dynamic critical exponent $z$ is independent of the path to the TCP.
	We obtain $z\nu \approx 1.003$ and $z\nu_{t} \approx 1.007$ from the fitting, leading to $z \approx 1.998$ or $z\approx 2.006$.
	Note that $z=2-\eta$ corresponds to a model with a non-conserved order parameter \cite{RevModPhys.49.435}, which is consistent with our model.

	\subsubsection{\texorpdfstring{$\chi{\rm SB}$}{TEXT} phase}
	In the $\chi{\rm SB}$ phase, the perturbations of $\theta$, $a_{t}$, and $a_{z}$ are coupled via the background solutions $\theta$ and $a_{t}$.
	However, we will take $\omega\to 0$ (and/or $k\to 0$) because we are interested only in the critical behaviors of the correlation length (and susceptibilities). 
	Then, the perturbation of $a_{z}$ is decoupled from the other perturbations.
	As a result, we solve the equations of motion for $\vartheta$ and ${\cal{A}}_{t}$, given by the pair of coupled differential equations.
	We numerically solve the coupled equations with proper boundary conditions, and compute the retarded Green's function by following the method in\,\cite{Kaminski:2009dh}.
	The details of the method are shown in appendix.
	
\begin{figure}[tbp]
		\centering 
		\includegraphics[width=7cm]{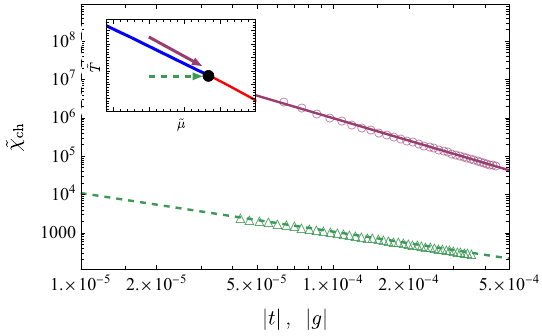}
		\includegraphics[width=7cm]{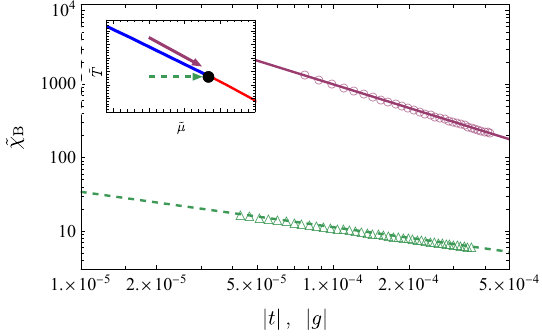}
		\caption{The critical behaviors of the chiral susceptibility (left panel) and the baryon number susceptibility (right panel) in log scale. 
			Each plot shows the critical behaviors for the two different paths denoted by the arrows in the inset.
		}
		\label{fig:gammaB}
	\end{figure}

	
	To determine the critical exponent $\gamma$ in the $\chi {\rm SB}$ phase, we compute the retarded Green's function $G^{R}_{ii}$ ($i=\vartheta, {\cal{A}}_{t}$) with $\omega \to 0$ and $k\to 0$.
	Figure \ref{fig:gammaB} shows the critical behaviors of the chiral susceptibility and the baryon number susceptibility.
	Here, we introduce $\chi_{\rm B}$ as the baryon number susceptibility which corresponds to $\chi_{2}$ in the previous sections. 
	By fitting the numerical data, we find
	\begin{align}
		\chi_{\rm ch} \propto     \left\{
		\begin{array}{ll}
			\abs{t}^{-\gamma} & (\text{along $\tau$-line})\\
			\abs{g}^{-\gamma_{t}} & (\text{along $t=0$})
		\end{array}
		\right., \hspace{0.5em}
		\chi_{\rm B} \propto     \left\{
		\begin{array}{ll}
			\abs{t}^{-\gamma_{2}} & (\text{along $\tau$-line})\\
			\abs{g}^{-\gamma_{2t}} & (\text{along $t=0$})
		\end{array}
		\right.,
		\label{eq:gammaB}
	\end{align}
	with $\gamma\approx1.948$, $\gamma_{t}\approx1.006$, $\gamma_{2}\approx1.068$, and $\gamma_{2t}\approx0.481$. 
	These values of the critical exponents agree with the mean-field values, i.e. (\ref{eq:chi1}), (\ref{eq:chi2}), and (\ref{eq:beta2}).
	
\begin{figure}[tbp]
		\centering 
		\includegraphics[width=7cm]{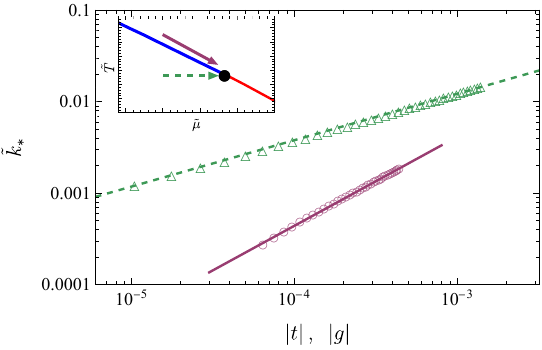}
		\includegraphics[width=7cm]{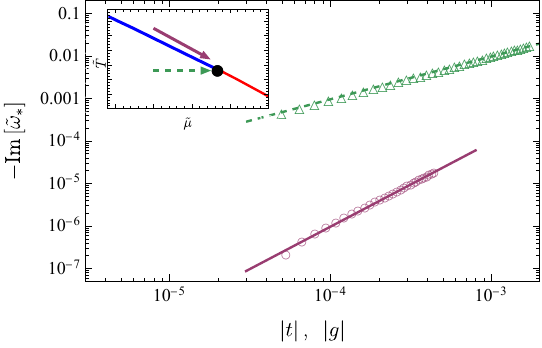}
		\caption{The left and right panel shows the critical behaviors of $k_{*}$ and $\omega_{*}$ in log scale, respectively. 
			Each plot shows the critical behaviors for the two different paths denoted by the arrows in the inset.
		}
		\label{fig:nuB}
	\end{figure}

	To determine the critical exponent $\nu$, we find the pole of the mixed Green's function and see the critical behavior of the correlation length by assuming the form of (\ref{eq:Green}).
	The left panel of Figure \ref{fig:nuB} shows the critical behavior of the pole $k_{*}$ with $\omega\to 0$.
	By fitting the numerical data, we find
	\begin{align}
		k_{*}^{-1}\sim\xi \propto     \left\{
		\begin{array}{ll}
			\abs{t}^{-\nu} & (\text{along $\tau$-line})\\
			\abs{g}^{-\nu_{t}} & (\text{along $t=0$})
		\end{array}
		\right.,
	\end{align}
	with $\nu\approx0.973$ and $\nu_{t}\approx0.507$, which also agree with the mean-field values.
	
	The dynamic critical exponent $z$ can be determined by finding the pole $\omega_{*}$ with $k\to 0$ as already discussed in the previous section.
	The right panel of Figure \ref{fig:nuB} shows the critical behavior of $-\Im \omega_{*}$ for the two different paths.
	By fitting the numerical data, we find $z\nu\approx1.989$ and $z\nu_{t}\approx1.007$ from (\ref{eq:znu}).
	Using $\nu\approx0.973$ and $\nu_{t}\approx0.507$, we obtain $z\approx2$ again.
	
\begin{figure}[tbp]
		\centering 
		\includegraphics[width=7cm]{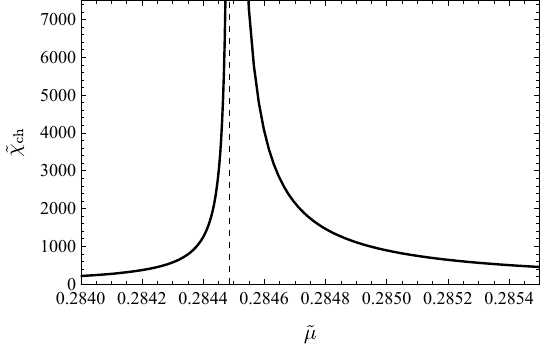}
		\includegraphics[width=7cm]{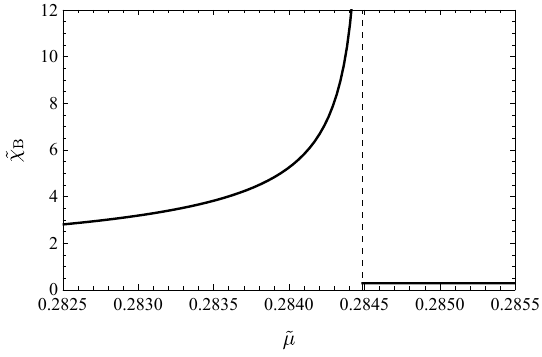}
		\caption{The critical behaviors of the chiral susceptibility (left panel) and the baryon number susceptibility (right panel) with $t=0$ fixed. The vertical dashed line denotes the critical value of the chemical potential $\tilde{\mu}=\tilde{\mu}_{t}$, corresponding to $g=0$.
		}
		\label{fig:sus}
	\end{figure}

	Before concluding this section, we show the critical behaviors of the chiral susceptibility and the baryon number susceptibility near the TCP.
	Figure \ref{fig:sus} shows the chiral susceptibility (left panel) and the baryon number susceptibility as a function of $\tilde{\mu}$ with $t=0$ fixed.
	As shown in the left panel of Figure \ref{fig:sus}, the chiral susceptibility is divergent in both the $\chi$SB and $\chi$SR phase.
	On the other hand, the baryon number susceptibility is divergent in the $\chi$SB phase whereas regular in the $\chi$SR phase as mentioned earlier.
	The critical exponents, $\gamma_{t}$ and $\gamma_{2t}$, associated with each singular behavior are given by (\ref{eq:gammaR}) and (\ref{eq:gammaB}).

	\subsubsection{Scaling hypothesis for correlation functions}
\begin{figure}[tbp]
		\centering 
		\includegraphics[width=7cm]{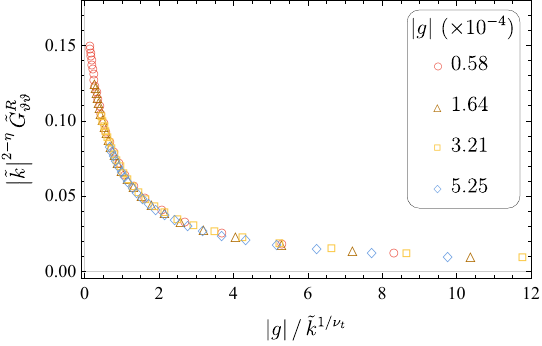}
		\includegraphics[width=7cm]{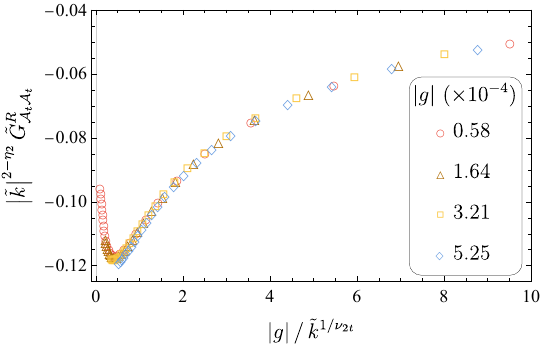}
		\caption{The left and right panel shows $|\tilde{k}|^{2-\eta}\tilde{G}^{R}_{\vartheta\vartheta}$ and $|\tilde{k}|^{2-\eta}\tilde{G}^{R}_{{\cal{A}}_{t}{\cal{A}}_{t}}$ with respect to $\abs{g}/\tilde{k}^{1/\nu_{t}}$, respectively. 
		}
		\label{fig:scaling1}
	\end{figure}

	Let us check the validity of the scaling hypothesis for the correlation functions in our system.
	Here, we again write the scaling hypothesis for the correlation functions:
	\begin{equation}
		G(t,g;k) \simeq \abs{k}^{\eta -2} {\cal{G}}^{(\pm)}\left( \frac{\abs{t}}{k^{1/\nu}}, \frac{\abs{g}}{k^{1/\nu_{t}}} \right).
		\label{eq:scaling3}
	\end{equation}
	To see this, we compute $\abs{k}^{2-\eta}\tilde{G}^{R}$ as a function of $\abs{g}/\tilde{k}^{1/\nu_{t}}$ for the path with $t=0$.
	Figure \ref{fig:scaling1} shows the numerical plots of $|\tilde{k}|^{2-\eta}\tilde{G}^{R}_{\vartheta\vartheta}$ (left panel) and $|\tilde{k}|^{2-\eta}\tilde{G}^{R}_{{\cal{A}}_{t}{\cal{A}}_{t}}$ (right panel) for several values of $g$.
	We set $\eta\approx0$ and $\nu_{t}\approx 0.5$ in the left panel, whereas $\eta_{2}\approx1.1$ and $\nu_{2t}\approx 0.52$ in the right panel.
	The coincidence of numerical data for several values of $\abs{g}$ implies that the correlation function satisfies the scaling hypothesis (\ref{eq:scaling3}) with these critical exponents.
	Here, we introduce the exponent $\nu_{2t}$ for the scaling hypothesis for $G^{R}_{{\cal{A}}_{t}{\cal{A}}_{t}}$.
	However, since the critical exponent $\nu_{t}$ is given by the critical behavior of the correlation length that is uniquely determined by the pole of the mixed correlation function $G_{ii}^{R}$, $\nu_{t}$ and $\nu_{2t}$ are expected to be identical for each scaling form.
	In addition, we find that the value of $\eta_{2}$ also agrees with the mean-field value, that is, (\ref{eq:nueta}).
	Together with other critical exponents, the critical exponents in the $\chi{\rm SB}$ phase satisfy the Fisher's scaling relations
	\begin{equation}
		\gamma_{t} = \nu_{t}\left(2-\eta \right), \hspace{0.5em} \gamma_{2t} = \nu_{t}\left(2-\eta_{2} \right).
	\end{equation}
	
\begin{figure}[tbp]
		\centering 
		\includegraphics[width=7cm]{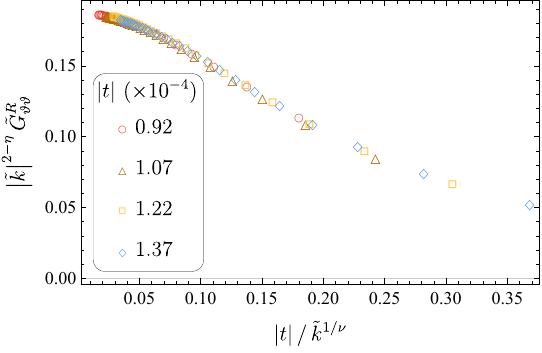}
		\includegraphics[width=7cm]{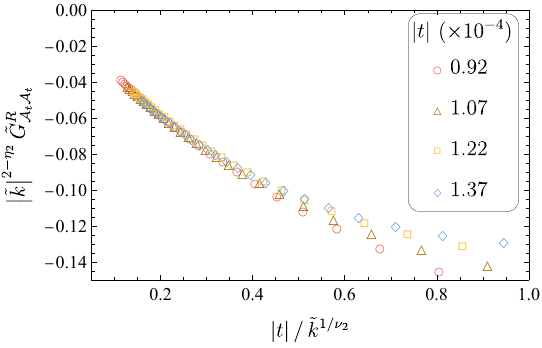}
		\caption{The left and right panel shows $|\tilde{k}|^{2-\eta}\tilde{G}^{R}_{\vartheta\vartheta}$ and $|\tilde{k}|^{2-\eta}\tilde{G}^{R}_{{\cal{A}}_{t}{\cal{A}}_{t}}$ with respect to $\abs{t}/\tilde{k}^{1/\nu}$, respectively. 
		}
		\label{fig:scaling2}
	\end{figure}

	We now see the dependence of $\abs{t}/k^{1/\nu}$ for the scaling function.
	To do so, we consider a path to the TCP along the triple line which is asymptotically $g=0$.
	Figure \ref{fig:scaling2} shows the numerical plots of $|\tilde{k}|^{2-\eta}\tilde{G}^{R}_{\vartheta\vartheta}$ (left panel) and $|\tilde{k}|^{2-\eta}\tilde{G}^{R}_{{\cal{A}}_{t}{\cal{A}}_{t}}$ (right panel) for several values of $t$ along the triple line.
	We set $\eta\approx0$ and $\nu\approx0.98$ in the left panel, whereas $\eta_{2}\approx0.90$ and $\nu_{2}\approx0.90$ in the right panel.
	The numerical data is coincident in the left panel, implying that the scaling hypothesis with the above critical exponents is satisfied again.
	In the right panel, one can see that the numerical data is also coincident for small $\abs{t}/\tilde{k}^{1/\nu_{2}}$.
	Note that we obtain $\nu \approx \nu_{2}$ again.
	In addition, we find that the critical exponents along the triple line satisfy the scaling relations:
	\begin{equation}
		\gamma=\nu(2-\eta), \hspace{0.5em} \gamma_{2}=\nu (2-\eta_{2}).
	\end{equation}
	In summary, we numerically confirm that the scaling hypothesis for the correlation functions (\ref{eq:scaling3}) is well justified with the critical exponents whose values agree with the mean-field values.
	
	\subsection{Critical end points}
\begin{figure}[tbp]
		\centering 
		\includegraphics[width=7cm]{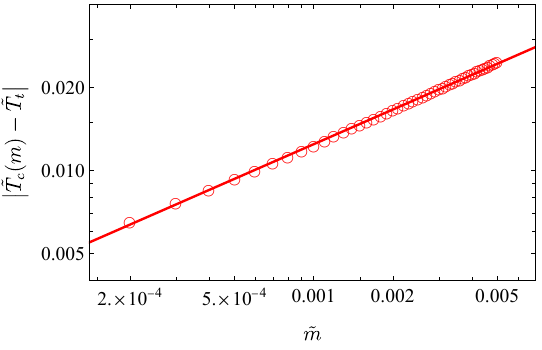}
		\includegraphics[width=7cm]{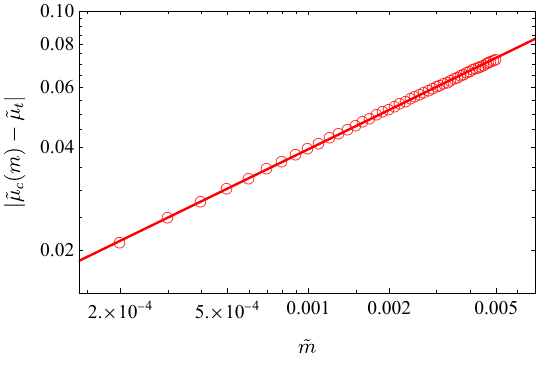}
		\caption{The critical behaviors of $| \tilde{T}_{c}(m)-\tilde{T}_{t} |$ (left panel) and $\abs{\tilde{\mu}_{c}(m)-\tilde{\mu}_{t}}$ (right panel) as a function of $\tilde{m}$ near the TCP.}
		\label{fig:CEP}
	\end{figure}

	In the last part of this section, we study the critical behavior of the critical end points.
	As seen in (\ref{eq:CEP1}) and (\ref{eq:CEP2}), the critical behavior of the critical end points are written as a function of the source $m$ near the TCP.
	Figure \ref{fig:CEP} shows the critical behaviors of $| \tilde{T}_{c}(m)-\tilde{T}_{t} |$ (left panel) and $\abs{\tilde{\mu}_{c}(m)-\tilde{\mu}_{t}}$ (right panel) near the TCP.
	By fitting the numerical data, we obtain
	\begin{equation}
		\abs{\tilde{T}_{c}(m)-\tilde{T}_{t}} \propto m^{0.417}, \hspace{0.5em} \abs{\tilde{\mu}_{c}(m)-\tilde{\mu}_{t}} \propto m^{0.382}, 
	\end{equation}
	which is roughly consistent with the expressions in (\ref{eq:CEP1}) and (\ref{eq:CEP2}), that is, they depend on $m^{2/5}$ for small $m$.
	
	\section{Conclusion and discussions} \label{sec:conclusion}
	In this paper, we study the critical phenomena at the TCP that emerges in the D3/D7 model.
	The TCP is associated with the chiral phase transition at finite temperature in the presence of the magnetic field and baryon number density according to the AdS/CFT correspondence.
	We numerically determine the values of the critical exponents ($\alpha, \beta, \gamma, \delta, \nu, \eta, z$) for two different paths to the TCP.
	One is a path which is asymptotically tangential to the phase boundary, such as the path along the triple line.
	The other is a path that is not asymptotically tangential to the phase boundary, such as the path with $T=T_{t}$ fixed.
	In addition, there are two different choices of parameters in the definition of the critical exponents: the order parameter and density.
	In summary, we define the 15 critical exponents ($\alpha, \alpha_{t}, \beta, \beta_{t}, \beta_{2}, \delta, \delta_{2}, \gamma, \gamma_{t}, \gamma_{2}, \nu, \nu_{t}, \eta, \eta_{2}, z$) in this paper.
	We confirm that all exponents determined in this paper take the mean-field values and satisfy the scaling relations between them, implying that the scaling hypothesis for the free energy and correlation functions hold.
	We directly check the scaling hypothesis for the correlation function by plotting the scaling function.
	Our results indicate that the critical phenomena at the TCP in the D3/D7 model are well described by the conventional Landau theory discussed in section \ref{sec:Landau}.
	We infer that the agreement with the mean-field like behaviors is due to the large-$N_{c}$ limit, where fluctuations are suppressed.
	
	Before concluding this paper, we have some remarks.
	As mentioned in section \ref{sec:intro}, the author studied the critical phenomena at the TCP that arise in the presence of an electric current in the D3/D7 model \cite{Matsumoto:2022nqu}.
	In \cite{Matsumoto:2022nqu}, we found that the critical phenomena at the TCP are asymmetric; the values of the critical exponents ($\gamma, \nu$) are different in the $\chi$SR and $\chi$SB phase.
	Thus, our results in the present paper imply that the tricritical phenomena in equilibrium are symmetric and show the mean-field like behaviors, whereas those in the non-equilibrium steady state regime are asymmetric and the critical exponents ($\gamma, \nu$) in the $\chi$SB phase are not the mean-field values.
	In other words, the study in this paper emphasizes that the tricritical phenomena at the current-driven TCP are characteristic of the non-equilibrium regime.
	It would be interesting to investigate the critical phenomena for a different path, for example along the triple line as studied in this paper, at the current-driven TCP and gain insight into the peculiar values of ($\gamma, \nu$).
	Another interesting direction is to analytically formulate the critical phenomena at the TCP. 
	Since we numerically confirm that the critical phenomena at the TCP show the mean-field like behaviors, they should be formulated by the Landau free energy functional form (\ref{eq:free}) with the coefficients expanded as (\ref{eq:a1}) and (\ref{eq:b1}).
	Hence, it would be interesting to investigate an effective theory derived from the DBI action.
	However, it is left for a future work.

	\acknowledgments
	The author thanks Shin Nakamura for the fruitful discussion.
	The author also thank RIKEN iTHEMS NEW working group for fruitful discussions.
	M.~M. is supported by National Natural Science Foundation of China with Grant No.~12047538.

	\appendix

	\section{Green's functions in the \texorpdfstring{$\chi{\rm SB}$}{TEXT} phase}
	In this appendix, we briefly review the method for calculating the Green's function in the system of coupled equations.
	The formalism basically follows the discussion in \cite{Kaminski:2009dh}.
	In this study, we consider the following perturbations of the background fields
	\begin{equation}
		\theta \to \theta(u)+ \delta \theta(t,z,u), \hspace{0.5em} A_{t}\to a_{t}(u)+ \delta a_{t}(t,z,u), \hspace{0.5em} A_{z}\to 0 + \delta a_{z}(t,z,u),
	\end{equation}
	where $\theta(u)$ and $a_{t}(u)$ are the background solutions. 
	Here, $\delta a_{z}$ is coupled because we assume that the perturbations depend on the spatial coordinate $z$.
	Expanding the action to the quadratic order, we obtain
	\begin{equation}
		S^{(2)} = -\frac{{\cal N}}{2} \int \dd t \dd ^{3}\vec{x} \dd u \left[ \left( \partial_{\alpha} \Phi^{T} \right) A^{\alpha\beta} \partial_{\beta} \Phi + \Phi^{T} B^{\alpha} \partial_{\alpha} \Phi + \Phi^{T}C\Phi\right],
		\label{eq:action21}
	\end{equation}
	where $\Phi = \left( \delta\theta, E_{L} \right)^{T}$ and $A^{\alpha\beta}$, $B^{\alpha}$, and $C$ are coefficient matrices.
	Here, we have used the gauge-invariant combination $E_{L} = -i(k \delta a_{t} +\omega \delta a_{z})$.
	We consider the Fourier transformation of the fields
	\begin{equation}
		\Phi(t,z,u) = \int\frac{\dd^{4} q}{(2\pi)^{4}} e^{-iqx}\tilde{\Phi}_{q}(u),
	\end{equation}
	where $q=(\omega,0,0,k)$ in our case. Inserting this into (\ref{eq:action21}), the quadratic action is given by
	\begin{equation}
		S^{(2)} = -\frac{{\cal N}}{2} \int \frac{\dd^{4} q}{(2\pi)^{4}} \int \dd u \left[ \tilde{\Phi}_{-q}^{T \prime} A_{q} \tilde{\Phi}_{q}' +\tilde{\Phi}^{T}_{-q} B_{q} \tilde{\Phi}_{q}' +\tilde{\Phi}^{T}_{-q} C_{q} \tilde{\Phi}_{q} \right],
		\label{eq:action22}
	\end{equation}
	where the prime denotes the derivative with respect to $u$.
	The coefficient matrices are explicitly given by
	\begin{align}
		A_{q} &= \frac{fu^{2}L_{0}}{DG}
		\begin{bmatrix}
			\omega^{2}-k^{2} f F &  -i k u^{4} f \theta' a_{t}' \\
			-i k u^{4} f \theta' a_{t}' & u^{2} H \\
		\end{bmatrix}
		, \\
		B_{q} &= -\frac{6fu^{2}\tan\theta L_{0}}{D}
		\begin{bmatrix}
			(\omega^{2}-k^{2} f)\theta' &  ik u^{2}a_{t}' \\
			0 & 0 \\
		\end{bmatrix}
		, \\
		C_{q} &= L_{0}
		\begin{bmatrix}
			3(-3+2\sec^{2}\theta)+\frac{u^{2}(\omega^{2}-k^{2}fF)}{fG}+\frac{9u^{4}\omega^{2}\tan^{2}\theta a_{t}'^{2}}{D} &  \frac{-ik u^{6}\theta' a_{t}'}{G} \\
			\frac{-ik u^{6}\theta' a_{t}'}{G} & -\frac{u^{4}H}{fG} \\
		\end{bmatrix}
		,
	\end{align}
	where 
	\begin{align}
		& F = 1-u^{4}a_{t}', \hspace{1em} G=1+u^{2}f\theta'^{2} -u^{4}a_{t}'^{2}, \hspace{1em} H = 1+u^{2}f \theta'^{2}, \nonumber \\
		& D = \omega^{2}H -k^{2}fG, \hspace{1em} L_{0}=\frac{\cos^{3}}{u^{5}}\sqrt{(1+B^{2}u^{4})G}, \nonumber
	\end{align}
	Here, we can find $A_{-q}=A_{q}^{*}$, $B_{-q}=B_{q}^{*}$, and $C_{-q}=C_{q}^{*}$.
	To avoid double counting, we consider the momentum integration only over positive momenta as follows:
	\begin{equation}
		S^{(2)} = \int \dd q_{>} \int \dd u \left[ \tilde{\Phi}_{-q}^{T \prime} \left(A_{q} + A_{q}^{\dagger}\right) \tilde{\Phi}_{q}' +\tilde{\Phi}^{T}_{-q} B_{q} \tilde{\Phi}_{q}' +\tilde{\Phi}^{T \prime}_{-q} B_{q}^{\dagger} \tilde{\Phi}_{q} +\tilde{\Phi}^{T}_{-q} \left( C_{q}+C_{q}^{\dagger} \right)\tilde{\Phi}_{q} \right],
		\label{eq:action23}
	\end{equation}
	where $\int \dd q_{>} = (2\pi)^{-4}\int_{0}^{\infty}\dd\omega \int_{\mathbb{R}^{3}} \dd \vec{k}$.
	For convenience, we introduce a new variable 
	\begin{equation}
		\tilde{\Phi}_{q}(u) = \varrho(u)\bar{\Phi}_{q},
	\end{equation}
	where $\varrho(u) ={\rm diag}(u,1)$. This factor makes the asymptotic behavior of the fields into $\bar{\Phi}_{q}(u\to 0) = \Phi_{q}^{(0)}$, where $\Phi_{q}^{(0)}$ can be interpreted as the source of the dual operator.
	Introducing the corresponding coefficient matrices by
	\begin{align}
		&\bar{A}_{q} = \varrho {A}_{q} \varrho, \\
		&\bar{B}_{q} = \varrho {B}_{q} \varrho + \varrho' {A}_{q} \varrho, \\
		&\bar{C}_{q} = \varrho {C}_{q} \varrho + \varrho {B}_{q} \varrho' + \varrho' {B}_{q}^{\dagger} \varrho + \varrho' {A}_{q} \varrho',
	\end{align}
	the quadratic action can be written in the same form as (\ref{eq:action22}).
	Hence, the equations of motion are given by
	\begin{equation}
		\left( \left(\bar{A}_{q}+\bar{A}_{q}^{\dagger}\right) \bar{\Phi}_{q}' +\bar{B}_{q}^{\dagger}\bar{\Phi}_{q} \right)' - \bar{B}_{q}\bar{\Phi}_{q}' - \left(\bar{C}_{q}+\bar{C}_{q}^{\dagger}\right)\bar{\Phi}_{q}=0.
		\label{eq:EOM}
	\end{equation}
	We now consider the solutions $\bar{\Phi}^{I}_{k}$ which are sources for operators ${\cal{O}}^{I}$, where $I=1,2$ in our case.
	A particular solution that sources ${\cal{O}}^{I}$ is given by a set of functions $\left\{ \bar{\Phi}_{k}^{I}(u) \right\}$ that satisfy $\bar{\Phi}_{k}^{I}(\varepsilon)=\delta^{I}_{~J}\varphi_{k}^{J}$, where $\varphi_{k}^{J}$ is the source of the corresponding operator ${\cal{O}}^{J}$ at the boundary $u=\varepsilon$. We have introduced the small cutoff $\varepsilon$ at the boundary.
	Since the fields are coupled, a set of functions $\left\{ \bar{\Phi}_{k}^{I}(u) \right\}$ source a linear combination of all the operators.
	Then, we define the solution matrix $F^{I}_{~J}$ as
	\begin{align}
		&\bar{\Phi}^{I}_{q}(u) = F^{I}_{~J}(q,u) \varphi_{q}^{J}, \\
		&\bar{\Phi}^{I}_{-q}(u) = F^{I}_{~J}(-q,u) \varphi_{-q}^{J} = \varphi_{-q}^{J} F_{\,J}^{\dagger \,I}(q,u), 
		\label{eq:Fmat}
	\end{align}
	where $F^{I}_{~J}(q,\varepsilon)=\delta^{I}_{~J}$ is satisfied by definition.
	Inserting (\ref{eq:Fmat}) into the action and using the equations of motion, the boundary action can be written by
	\begin{align}
		S^{(2)}_{\rm bdry} = \left. \int \dd k_{>} \,\varphi^{I}_{-q} {\cal{F}}_{IJ}(q,u)\varphi_{q}^{J} \right|_{\varepsilon}^{u_{\rm H}},
	\end{align}
	where ${\cal{F}}(q,u) = F^{\dagger} \left( A+A^{\dagger} \right)F' +F^{\dagger}B F$.
	According to the holographic prescription \cite{Son:2002sd}, we obtain the general form of the retarded Green's function
	\begin{equation}
		G^{R}_{IJ}(q) = - \lim_{\varepsilon \to 0} {\cal{F}}_{IJ}(q,\varepsilon).
		\label{eq:flux}
	\end{equation}
	
	In general, it is difficult to find an analytic solution to the coupled equations (\ref{eq:EOM}).
	Hence, we need to employ a numerical method to solve the equations with appropriate boundary conditions.
	For our purpose, we impose the in-going wave boundary condition at the horizon:
	\begin{equation}
		\bar{\Phi}(u) = \left(u_{\rm H} - u \right)^{-\frac{i \omega}{4\pi T}} \bar{\Phi}_{\rm reg}(u),
	\end{equation}
	where $\bar{\Phi}_{\rm reg}$ is the regular function at the horizon.
	To calculate the retarded Green's function numerically, we first consider a set of linearly independent and in-going wave solutions $\left\{ \phi_{(1)},\phi_{(2)} \right\}$.
	In our calculation, we set $\phi_{(J)}^{I}(u_{\rm H})=\delta^{I}_{~J}$ with $I,J\in \{1,2\}$.
	Solving the equations of motion numerically with these boundary conditions at the horizon, we can construct a matrix of these solutions:
	\begin{equation}
		H^{I}_{~J}(q,u) = \phi^{I}_{(J)}(u),
	\end{equation}
	where $H^{I}_{~J}(q,u_{\rm H}) = \delta^{I}_{~J}$ by definition.
	The solution matrix $F(q,u)$ that is normalized at the boundary is linearly related to $H(q,u)$ by
	\begin{equation}
		F(q,u) = H(q,u) \cdot H^{-1}(q,\varepsilon).
	\end{equation}
	Near the boundary, the matrix $H$ is asymptotically given by
	\begin{equation}
		H^{I}_{~J}(q,u\to 0) = \mathbb{S}^{I}_{~J}(q) + \mathbb{O}^{I}_{~J} (q) u^{\Delta^{I}} + \cdots,
	\end{equation}
	where $\mathbb{S}$ and $\mathbb{O}$ are the coefficient matrices in a power series of $H$ at the boundary, and $\Delta^{I}$ is the corresponding power of each normalizable mode.
	Inserting this into (\ref{eq:flux}), the retarded Green's function is given by 
	\begin{align}
		G^{R}(q) = - \lim_{\varepsilon \to 0} {\cal{F}}(q,\varepsilon) &= - \lim_{\varepsilon \to 0}  \left[ \left(A+A^{\dagger} \right)F' +B^{\dagger}  \right]_{u=\varepsilon} \nonumber \\
		& = - \lim_{\varepsilon \to 0}  \left[ \Delta^{I} \varepsilon^{\Delta^{I}-1} \left(A+A^{\dagger} \right)\mathbb{O}\cdot \mathbb{S}^{-1} +B^{\dagger}  \right]_{u=\varepsilon},
	\end{align}
	where we have used $F(q,\varepsilon)=1$ in the second equality.
	Note that the Green's function is obviously ill-defined if $\det \mathbb{S}=0$. 
	In other words, the poles of the Green's function is determined by imposing the condition
	\begin{equation}
		\det H(q,\varepsilon)=0.
	\end{equation}
	These poles corresponds to the quasi-normal modes in the gravity picture.
	The quasi-normal modes for the massless case in the D3/D7 model have been studied in \cite{Hoyos-Badajoz:2006dzi,Kaminski:2009ce,Atashi:2022ufl}.
	
	
	\bibliographystyle{JHEP}
	\bibliography{main}
	
\end{document}